\documentclass[12pt]{article}
\usepackage{amsmath,amssymb,amsfonts,epsfig,graphicx}
\newcommand{\be}{\begin{equation}}
\newcommand{\ee}{\end{equation}}

\newcommand{\bse}{\begin{subequations}}
\newcommand{\ese}{\end{subequations}}
\newcommand{\bea}{\begin{eqnarray}}
\newcommand{\eea}{\end{eqnarray}}
\newcommand{\ba}{\begin{array}}
\newcommand{\ea}{\end{array}}

\def\bp{\bar{\Pi}}

\def\half{\frac{1}{2}}

\makeatletter \@addtoreset{equation}{section}

\makeatother
\begin{document}
\baselineskip 18pt%

\begin{titlepage}
\vspace*{1mm}%
\hfill%
\vbox{
    \halign{#\hfil \cr
\;\;\;\;\;\;\;\;\;\; IPM/P-2007/039 \cr
           \;\;\;\;\;\;\;\;\;\;\;\;\;\; SUT-P-07-2a    \cr
arXiv:0708.2058 {\tt [hep-th]} \cr
           \cr
           } 
      }  
\vspace*{10mm}%

\centerline{{\Large {\bf  \textsl{Electrified BPS Giants:}}}}%
\vspace*{3mm}
\centerline{{\large {\bf BPS configurations on Giant Gravitons with Static Electric Field }}}%
\vspace*{7mm}
\begin{center}
{\bf \large{M. Ali-Akbari$^{1,2}$, M. M. Sheikh-Jabbari$^{1}$}}
\end{center}
\begin{center}
\vspace*{0.4cm}
{\it {$^1$Institute for Studies in Theoretical Physics and Mathematics (IPM)\\
P.O.Box 19395-5531, Tehran, IRAN\\
$^2$Department of Physics, Sharif University of Technology\\
P.O.Box 11365-9161, Tehran, IRAN}}\\
{E-mails: {\tt aliakbari, jabbari @theory.ipm.ac.ir}}%
\vspace*{1.5cm}
\end{center}

\begin{center}{\bf Abstract}\end{center}
\begin{quote}
We consider  D3-brane action in the maximally supersymmetric type
IIB plane-wave background. Upon  fixing the light-cone gauge, we
obtain the light-cone Hamiltonian which is manifestly
supersymmetric. The 1/2 BPS solutions of this theory (solutions
which preserve 16 supercharges) are either of the form of spherical
three branes, the giant gravitons, or zero size point like branes.
We then construct specific classes of 1/4  BPS solutions of this
theory in which static electric  field on the brane is turned on.
These solutions are  deformations about either of the two 1/2 BPS
solutions. In particular, we study in some detail 1/4 BPS
configurations with electric dipole on the three sphere giant,
\emph{i.e.} BIons on the giant gravitons, which we hence  call
BIGGons. We also study BPS configurations corresponding to turning
on a background uniform constant electric  field. As a result of
this background electric field the three sphere giant is deformed to
squashed sphere, while the zero size point like branes turn into
circular or straight fundamental strings in the plane-wave
background, with their tension equal to the background electric
field.

\end{quote}
\end{titlepage}
%
%
\tableofcontents

\section{Introduction}

The idea that a non-perturbative formulation of string theory
dynamics can be obtained from (some particular limits of) D-brane
dynamics  has proved very fruitful; the BFSS matrix model
\cite{BFSS} and the AdS/CFT duality \cite{AdS/CFT} are indeed
outcomes of this viewpoint. To exploit  this idea further one needs
to have a much better grasp of the brane dynamics than what we have
now.

One of the first steps in this direction was taken in
\cite{Callan-Maldacena} trying to reproduce the Polchinski's picture
for D-branes \cite{Polchinski}, namely open strings with Dirichlet
boundary conditions ending on the brane, from the D-brane theory.
The low energy effective field theory of a single D$p$-brane is a
$p+1$ supersymmetric $U(1)$ gauge theory with 16 supercharges which
is described by the Dirac-Born-Infeld action plus the Chern-Simons
terms through which brane couples to background RR form fields. At
first order in $\alpha'$, \emph{i.e.} at the quadratic action level,
the theory reduces to supersymmetric $p+1$ dimensional  Maxwell
theory. In the supersymmetric Maxwell limit, it was shown that a BPS
configuration corresponding to an electric charge on the $p$
dimensional brane, which was called a ``spike'',  correctly
reproduces the behavior one expects from the open strings ending on
branes \cite{Callan-Maldacena}. This analysis was also extended to
the full Born-Infeld action and the ``spikes'' were hence called
BIons \cite{Callan-Maldacena, Gibbons}.

The above mentioned analysis was mainly performed for D-branes
residing in a background flat space. The $AdS_5\times S^5$ and the
ten dimensional plane-wave background, which can be obtained from
the former by taking the Penrose limit \cite{BMN}, are of special
interest both because they are the only maximally supersymmetric
type IIB backgrounds and more importantly because of the AdS/CFT
duality. The $AdS_5\times S^5$ background allows  1/2 BPS spherical
D3-brane configurations, the giant gravitons \cite{Giant}. It was
also shown that the plane-wave background also admits similar 1/2
BPS spherical 3-brane configurations e.g. see \cite{Takayanagi}.

Analyzing BPS states, due to protection by supersymmetry, has been
one of the most instructive ways in understanding and checking the
AdS/CFT or plane-wave/CFT duality. Among these BPS states many of
them could be related to deformations of a 1/2 BPS giant graviton
state. These deformations can come in two classes, one in which the
spherical shape of the giant has been deformed (e.g. see
\cite{Mikhailov}) or those which besides the shape we have also
turned on an electromagnetic gauge field on the brane (e.g. see
\cite{Electric-Giants, Gianthedge,squashed-giants}). In this paper
we focus on the analysis of the second class and in particular the
cases involving the static electric  fields on the giant and
classify all 1/4  BPS configurations for a given static electric
field.

This paper is organized as follows. In section 2, we start with
the D3-brane action on the plane-wave background. Fixing the
light-cone gauge and the $\kappa$-symmetry we obtain the full
supersymmetric light-cone Hamiltonian of the D3-brane in the
plane-wave background. We also present the complete superalgebra
of the Hamiltonian. In section 3, we study several 1/4 BPS
configurations involving various static electric fields. In the
absence of background electromagnetic fields, we have 1/2 BPS
configuration which are either of the form of finite size
spherical 3-brane, the giant gravitons, or a point like 3-brane, a
spherical brane of zero size. In particular we study the case
where the electric field is sourced by an electric dipole on the
three sphere giant, i.e. the counterpart of BIons on the Giant
Gravitons, which will hence be called BIGGons.  This generalizes
the Giant Hedgehogs of \cite{Gianthedge} to the full Born-Infeld
theory. We show that the BIGGons, unlike their flat brane
counterparts, have a finite extent and are not stretched to
infinity. We also study configurations with constant electric
field, showing that the electric field, similarly to the magnetic
field \cite{squashed-giants}, deforms (squashes) the three sphere
giant. As a result of background electric field on the point like
brane configurations, the brane behaves as a fundamental string on
the plane-wave background with tension equal to the electric
field. Section 4 contains our concluding remarks and outlook. Four
Appendices are added to fix our fermionic notation, introduce the
``Polyakov form'' of the DBI action, present some details of
light-cone gauge fixing and analysis of BPS equation.
%
%
\section{D3-brane light-cone Hamiltonian in the plane-wave background and its supersymmetry algebra}

In this section we work out  explicit form of the D3-brane
light-cone Hamiltonian on the maximally supersymmetric type IIB
plane-wave background. The plane-wave geometry is given by
\bse\label{planewave-bg}
\begin{align}
 ds^2 &=-2dX^+dX^--\mu^2(X^iX^i+X^aX^a)(dX^+)^2+dX^idX^i+dX^adX^a\\
C_{+ijk}&=-\frac{\mu}{g_s}\epsilon_{ijkl}X^l\ \ ,\quad
C_{+abc}=-\frac{\mu}{g_s}\epsilon_{abcd}X^d\\ e^\phi&=g_s=constant
\end{align}\ese%
where $i,a=1,2,3,4$ and $C$ is the four-form potential of the
self-dual five-form of type IIB. We have chosen our coordinates to
make manifest the $SO(4)\times SO(4)$ symmetry of the transverse
directions labeling vectors of the two $SO(4)$'s with $i,a$, as well
as the translation symmetry in $X^+$ and $X^-$ directions. In the
above metric $\frac{\partial}{\partial X^-}$ is a (globally defined)
null Killing direction and $\frac{\partial}{\partial X^+}$ is time
like.
 For a more detailed discussion on the isometries of the background we
refer the reader to \cite{0310119}.
%
%
\subsection{Supersymmetric D3-brane action in the plane-wave
background}%

The low energy supersymmetric effective action for a D3-brane in the
plane-wave background is \cite{0211178} %
\be\label{DBI} S=-\frac{1}{g_{s}}\int
d^{4}\sigma\sqrt{-det\big(g_{\hat{\mu}\hat{\nu}}+F_{\hat{\mu}\hat{\nu}}+M_{\hat{\mu}\hat{\nu}}\big)}+\int
{\cal L}_{WZ}\ee%
where%
\bse\label{pull-backs}
\begin{align}
 g_{\hat{\mu}\hat{\nu}}&=-2\partial_{\hat{\mu}}X^+\partial_{\hat{\nu}}X^-
 -\mu^2(X^I)^2\partial_{\hat{\mu}}X^+\partial_{\hat{\nu}}X^++\partial_{\hat{\mu}}X^I\partial_{\hat{\nu}}X^I\\
 M_{\hat{\mu}\hat{\nu}}&=2i\partial_{(\hat{\mu}}X^{+}(\bar{\psi}\bar{\gamma}^-\partial_{\hat{\nu})}\psi
 +\psi\bar{\gamma}^-\partial_{\hat{\nu})}\bar{\psi})-4\mu\bar{\psi}\bar{\gamma}^-\Pi\psi\partial_{\hat{\mu}}X^+
 \partial_{\hat{\nu}}X^+\\
 F_{\hat{\mu}\hat{\nu}}&=\partial_{\hat{\mu}}A_{\hat{\nu}}-\partial_{\hat{\nu}}A_{\hat{\mu}}
 -2i\partial_{[\hat{\mu}}X^{+}(\psi\bar{\gamma}^-\partial_{\hat{\nu}]}\psi
 +\bar{\psi}\bar{\gamma}^-\partial_{\hat{\nu}]}\bar{\psi})
\end{align}
\ese%
The hatted Greek indices are used for the worldvolume coordinate
ranging over $0,1,2,3$. The capital Latin indices, $I,J,\cdots$ are
used to denote the eight transverse directions, that is $I=(i,a)$,
in particular, $X^I$ denote the eight transverse embedding
coordinates of the brane. The $A_{\hat{\mu}}$ is the $U(1)$ gauge
field on the brane. Here we have set $2\pi\alpha'=1$ and powers of
$\alpha'$ can be recovered once needed, through dimensional
analysis. The parenthesis in the expression for
$M_{\hat{\mu}\hat{\nu}}$ means symmetrization on indices and hence
$M_{\hat{\mu}\hat{\nu}}$ is symmetric,
$M_{\hat{\mu}\hat{\nu}}=M_{\hat{\nu}\hat{\mu}}$. $\psi$'s are the
sixteen component complex but chiral fermions of type IIB. Note that
 the expressions in \eqref{pull-backs} have been written after
fixing the $\kappa$-symmetry as \cite{0211178}
\be\label{kappa-sym}%
{\bar\gamma}^+\psi={\bar\gamma}^+\bar\psi=0. \ee%
In this part we are employing the fermionic conventions of
\cite{0211178} which we have summarized in Appendix
\ref{Metsaev-fermion}. It is also useful to note that the last term
in $M$ which is linear in $\mu$ is coming from the coupling of
fermions to the background self-dual five-form flux.
Finally the Wess-Zumino part, after fixing the $\kappa$-symmetry as in \eqref{kappa-sym}, is%
\be \begin{split}\label{WZ-action}%
 {\cal
 L}_{WZ}=&-\epsilon^{\hat{\theta}\hat{\mu}\hat{\nu}\hat{\rho}}\partial_{\hat{\theta}}X^+
 \big[\partial_{\hat{\mu}}X^I
 \partial_{\hat{\nu}}X^I\bar{\psi}\gamma^{-IJ}\partial_{\hat{\rho}}\psi+\frac{1}{2}F_{\hat{\mu}\hat{\nu}}
 (\psi\bar{\gamma}^-\partial_{\hat{\rho}}\psi
 -\bar{\psi}\bar{\gamma}^-\partial_{\hat{\rho}}\bar{\psi})\big]\cr
 +&\ \frac{\mu}{6}
 \epsilon^{\hat{\theta}\hat{\mu}\hat{\nu}\hat{\rho}}\partial_{\hat{\theta}}X^+\big[\epsilon^{ijkl}
 X^i\partial_{\hat{\mu}}X^j\partial_{\hat{\nu}}X^k\partial_{\hat{\rho}}X^l+\epsilon^{abcd}
 X^a\partial_{\hat{\mu}}X^b\partial_{\hat{\nu}}X^c\partial_{\hat{\rho}}X^d\big].
\end{split}\ee%
 One can check that the Born-Infeld part
and the Wess-Zumino part of the action are individually
supersymmetric.

For the later use we separate the symmetric and anti-symmetric parts
of the matrix under the square-root: %
\be\label{N} N_{\hat{\mu}\hat{\nu}}\equiv (g_{\hat{\mu}\hat{\nu}}+
M_{\hat{\mu}\hat{\nu}})+F_{\hat{\mu}\hat{\nu}}
\ee%
and denote its inverse matrix by $N^{\hat{\mu}\hat{\nu}}$. The
symmetric and anti-symmetric parts of $N^{\hat{\mu}\hat{\nu}}$
respectively denoted by $G^{\hat{\mu}\hat{\nu}}$ and
$\theta^{\hat{\mu}\hat{\nu}}$ have the interpretation of
(supersymmetric) \emph{open string metric} and the
\emph{noncommutativity parameter} \cite{S-W}. In what follows we
denote the inverse of open string metric by
$G_{\hat{\mu}\hat{\nu}}$.

\subsection{Fixing the light-cone gauge, the light-cone Hamiltonian}
The supersymmetric brane action enjoys three class of local gauge
symmetries, the area preserving diffeomorphism (APD) invariance on
the worldvolume, its fermionic counterpart the $\kappa$-symmetry and
the $U(1)$ gauge symmetry. In the light-cone gauge  we fix a part of
the APD's, those which mix worldvolume time and spatial coordinates
while the spatial APD's are still un-fixed. We fix the fermionic
$\kappa$-symmetry completely by throwing away half of the
un-physical original 32 fermions. This latter we have done by
imposing \eqref{kappa-sym}. In the terminology of constrained
systems, these gauge fixing conditions are primary constraints and
one should make sure the consistency of these constraints, which in
turn lead to a set of secondary constraints. This work for the
supersymmetric D-brane action, in a general (not necessarily
light-cone gauge), but in the flat space background has been
performed in \cite{Bergshoeff, Kamimura}.

To fix the light-cone gauge we separate the space and time indices
on the brane worldvolume as
$\sigma^{\hat{\mu}}=(\tau=\sigma^0,\sigma^r),\ r=1,2,3$ the
space indices. The light-cone gauge is fixed by choosing%
\be\label{lightcone-time}
X^+=\tau\ . \ee%
To ensure that the above solution of $X^+$ is maintained by the
dynamics we use the time-space mixing part of the APD's and set the
time-space components of the open string metric equal to zero
\footnote{This equation parallels the ``level-matching condition''
in string theory. One way to see this is  to write the DBI action in
the ``Polyakov form'' by introducing an auxiliary worldvolume
metric, $h_{\hat{\mu}\hat{\nu}}$ (see Appendix \ref{Polyakov}). In
this language the light-cone gauge fixing amounts to setting
$h_{0r}=0$. In order this choice to be respected by the dynamics one
should make sure that the equation of motion for $h_{0r}$ is
satisfied, that is we demand
\[\frac{\delta {\cal L}}{\delta h_{0r}}=\frac{\partial {\cal L}}{\partial
h_{0r}}=0,\] (or impose it as a constraint). From the ``Polyakov
form'' of DBI action \eqref{polyakov-action} it is immediately seen
that $\frac{\partial
{\cal L}}{\partial h_{0r}}=G^{0r}$.}, {\it i.e.}%
 \be\label{Level-match}
N^{0r}+N^{r0}\equiv G^{0r}=G_{0r}=\big((g+M)-F(g+M)^{-1}F\big)_{0r}=0.%
\ee%
The above is the generalization of level matching condition for
strings to the D3-brane case. It is notable that for the plane-wave
case,  \eqref{Level-match} is independent of $\mu$, \emph{i.e.} it
has the same form as in the flat space background. The ``quantum
version'' of this fact has also been manifested in the Gauss law of
the tiny graviton matrix theory \cite{TGMT} as well as in the
analysis of \cite{Torabian} in which the matrix theory formulation
of type IIB string theory is obtained from non-BPS D0-branes.  The
above can be used to identify $\partial_r X^-$ in terms of other
dynamical variables and hence using \eqref{Level-match} both
$X^{\pm}$ are completely removed from the light-cone dynamics.

 In the
light-cone gauge, $M_{rs}=0$  while $M_{0r}$ and $M_{00}$ are
non-zero. Moreover, in the light-cone gauge the term in the WZ
action which is proportional to the gauge field $F$  becomes a total
derivative and hence can be dropped.

Next we note that in the plane-wave background, $X^-$ and $X^+$ are
cyclic variables and hence the corresponding conjugate momenta
respectively%
\be\label{p+Hlc}%
 p^+=\frac{\partial{\cal
L}}{\partial(\partial_{\tau}X^-)}=\frac{1}{g_{s}}\sqrt{-detN}N^{00},
\qquad H_{lc} \equiv P^-= \frac{\partial{\cal
L}}{\partial(\partial_{\tau}X^+)},\ee%
 ($H_{lc}$ is the light-cone
Hamiltonian density) are constants of motion.

Using properties of the determinant and some matrix identities, one
can eliminate the $\partial_\tau X^-$ dependence in the light-cone
Hamiltonian \cite{Gianthedge} and after some algebraic manipulations
(some of which have been gathered in the Appendix
\ref{light-cone-Appendix}) the light-cone
Hamiltonian is obtained to be%
\be\label{Hlc}\begin{split} {\cal H}=& \int
d^3\sigma\biggl\{\frac{(P^I)^2}{2p^+}+\frac{(P^I_E)^2}{2p^+}+\frac{1}{2}\mu^2p^+(X^I)^2+\frac{1}{2\cdot
3!p^+g_s^2}\{X^I,X^J,X^K\}^2+ \frac{1}{2p^+g_s^2}(B^I)^2\cr
+&\frac{\mu}{6g_s}\bigg(\epsilon^{ijkl}X^i\{X^j,X^k,X^l\}+\epsilon^{abcd}X^a\{X^b,X^c,X^d\}\bigg)\\%
+&\mu\psi^{\dagger\alpha\beta}\psi_{\alpha\beta}+\frac{2}{p^+g_s}\bigg(\psi^{\dagger\alpha\beta}(\sigma^{ij})_\alpha^{\
\delta}\{X^i,X^j,\psi_{\delta\beta}\}+\psi^{\dagger\alpha\beta}(\sigma^{ab})_\alpha^{\
\delta}\{X^a,X^b,\psi_{\delta\beta}\}\bigg)\cr
+&\mu\psi^{\dagger\dot{\alpha}\dot{\beta}}\psi_{\dot{\alpha}\dot{\beta}}+\frac{2}{p^+g_s}\bigg(\psi^{\dagger\dot{\alpha}\dot{\beta}}(\sigma^{ij})_{\dot{\alpha}}^{\
\dot{\delta}}\{X^i,X^j,\psi_{\dot{\delta}\dot{\beta}}\}+\psi^{\dagger\dot{\alpha}\dot{\beta}}(\sigma^{ab})_{\dot{\alpha}}^{\
\dot{\delta}}\{X^a,X^b,\psi_{\dot{\delta}\dot{\beta}}\}\bigg)\biggr\}
\end{split}\ee%
 In the above we have used
$SO(4)\times SO(4)$ representation for the fermions (see Appendix
\ref{fermion-so4so4}), $P^I$ and $P^I_E$ are respectively momenta
conjugate to $X^I$ and the gauge field $A_r$ times $\partial_r X^I$, \eqref{PI-def}, \eqref{PrE-def} and \eqref{PIE}, and%
\be\label{BI}
B^I=B^r\partial_rX^I=-\frac{1}{\sqrt{2}}\epsilon^{rsp}F_{sp}\partial_rX^I
\ .%
\ee%
Finally the brackets are Nambu 3-brackets defined as (e.g. see
\cite{TGMT})%
\be\label{N3B}%
\{F, G, K\}=\epsilon^{rps}\partial_r F\partial_s G\partial_s K.%
\ee%

The above Hamiltonian should be supplemented by the secondary
constraint coming from the $U(1)$ gauge symmetry:%
\be\label{Gauss-law}
\partial_r P^r_E=0.
\ee%
Noting the results of \cite{Kamimura} it can be shown that fixing
the light-cone gauge by imposing \eqref{kappa-sym} and
\eqref{lightcone-time}, leads to no further secondary constraints.
However, one should still make sure that the physical configurations
of the above Hamiltonian is satisfying  \eqref{Level-match} which
can be simplified to%
\be\label{Level-match-LC}%
p^+\partial_rX^-=P^I\partial_r X^I+\bar\psi
\bar\gamma^-\partial_r\psi+
\psi\bar\gamma^-\partial_r\bar\psi+F_{rs}P^s_E.%
\ee

The light-cone Hamiltonian is invariant under local three
dimensional APD's and also the $U(1)$ gauge symmetry (which can be
fixed in any gauge, the light-cone gauge or otherwise). It has also
global symmetries, such as $psu(2|2)\times psu(2|2)\times u(1)_H$
superalgebra, which will be made explicit in the next subsection,
the $\mathbb{Z}_2$ symmetry which exchanges $X^i$ and $X^a$
directions (or identically exchanges the two $psu(2|2)$ factors of
the superalgebra) and the electric-magnetic duality %
\be\label{duality}%
 P_E^I\longleftrightarrow \frac{B^I}{g_s}.
\ee
%
%
\subsection{The light-cone supersymmetry algebra}

As mentioned the light-cone Hamiltonian is invariant under the
dynamical part of the plane-wave superalgebra which is
$psu(2|2)\times psu(2|2)\times u(1)_H$. It happens that the relevant
superalgebra to our case is in fact the ``extended'' $psu(2|2)\times
psu(2|2)\times u(1)_H$ superalgebra \cite{TGMT-SUSY-extension}:

$\bullet$ \emph{The fermionic anti-commutators}
\bse\label{susy-dynamical}\begin{align}%
\{Q_{\dot{\alpha}\beta},Q^{\dagger\dot{\rho}\lambda}\}&=\delta^{\dot{\rho}}_{\dot{\alpha}}\delta^\lambda_\beta
{\cal H}
+\frac{\mu}{2}(i\sigma^{ij})^{\dot{\rho}}_{\dot{\alpha}}\delta^\lambda_\beta
{\cal
J}_{ij}+\frac{\mu}{2}\delta^{\dot{\rho}}_{\dot{\alpha}}(i\sigma^{ab})^\lambda_\beta
{\cal
J}_{ab}+(i\sigma^{ij})^{\dot{\rho}}_{\dot{\alpha}}(i\sigma^{ab})^\lambda_\beta
{\cal R}_{ijab}\\
\{Q_{\alpha\dot{\beta}},Q^{\dagger\rho\dot{\lambda}}\}&=\delta^\rho_\alpha\delta^{\dot{\lambda}}_{\dot{\beta}}
{\cal
H}+\frac{\mu}{2}(i\sigma^{ij})^\rho_\alpha\delta^{\dot{\lambda}}_{\dot{\beta}}
{\cal
J}_{ij}+\frac{\mu}{2}\delta^\rho_\alpha(i\sigma^{ab})^{\dot{\lambda}}_{\dot{\beta}}
{\cal
J}_{ab}+(i\sigma^{ij})^{\rho}_{\alpha}(i\sigma^{ab})^{\dot{\lambda}}_{\dot{\beta}}
{\cal R}_{ijab}\end{align}\ese%
 where ${\cal H}$ is the light-cone Hamiltonian and ${\cal J}_{ij},\
 {\cal J}_{ab}$ are generators of the two $SO(4)$'s.
\bse\label{susyqq}\begin{align}
\{Q_{\dot{\alpha}\beta},Q_{\dot{\rho}\lambda}\}=& {\cal
Z}\epsilon_{\dot{\alpha}\dot{\rho}}\epsilon_{\beta\lambda}+ {\cal
Z}_{ijab}(i\sigma^{ij})_{\dot{\alpha}\dot{\rho}}(i\sigma^{ab})_{\beta\lambda}\\
\{Q_{\alpha\dot{\beta}},Q_{\rho\dot{\lambda}}\}=&{\cal
Z}\epsilon_{\alpha\rho}\epsilon_{\dot{\beta}\dot{\lambda}}+{\cal
Z}_{ijab}(i\sigma^{ij})_{\alpha\rho}(i\sigma^{ab})_{\dot{\beta}\dot{\lambda}}
\end{align}\ese%

$\bullet$ \emph{The fermionic-bosonic commutators}
\bse\begin{align}%
[{\cal
J}_{ij},Q_{\alpha\dot{\beta}}]=\frac{1}{2}(i\sigma^{ij})_\alpha^\rho
Q_{\rho\dot{\beta}}\ \ \ \ \ &,\ \ \ \ \ [{\cal
J}_{ij},Q_{\dot{\alpha}\beta}]=\frac{1}{2}(i\sigma^{ij})_{\dot{\alpha}}^{\dot{\rho}}
Q_{\dot{\rho}\beta}\\
[{\cal
J}_{ab},Q_{\alpha\dot{\beta}}]=\frac{1}{2}(i\sigma^{ab})_{\dot{\beta}}^{\dot{\rho}}
Q_{\alpha\dot{\rho}}\ \ \ \ \ &,\ \ \ \ \ [{\cal
J}_{ab},Q_{\dot{\alpha}\beta}]=\frac{1}{2}(i\sigma^{ab})_\beta^\rho
Q_{\dot{\alpha}\rho}\end{align}\ese%
\be [{\cal H},Q_{\alpha\dot{\beta}}]=0\ \ \ \ \ ,\ \ \ \ \
[{\cal H},Q_{\dot{\alpha}\beta}]=0\ee%
Note that extensions $ {\cal R}_{ijab},\ {\cal Z}_{ijab}$ are not
central because they do not commute with ${\cal J}$'s.

It is straightforward, but involves lengthy computations, to show
that this superalgebra  has an explicit realization in terms of the
D3-brane fields as:%
\be\label{Q}\begin{split}
 Q_{\dot{\alpha}\beta}=\frac{1}{\sqrt{2p^+}}\int d^3\sigma
 \bigg[&(P^i-i\tilde{X}^i)(\sigma^i)_{\dot{\alpha}}^{\ \ \rho}
 \psi_{\rho\beta}+(\frac{B^i}{g_s}+iP_E^i)(\sigma^i)^{\rho}_{\ \
 \dot{\alpha}}\psi^{\dagger}_{\rho\beta}\cr
 +&(P^a-i\tilde{X}^a)(\sigma^i)_{\beta}^{\ \ \dot{\rho}}
 \psi_{\dot{\alpha}\dot{\rho}}+(\frac{B^a}{g_s}+iP_E^a)(\sigma^a)^{\dot{\rho}}_{\
 \ \beta}\psi^{\dagger}_{\dot{\alpha}\dot{\rho}}\cr
 -&\frac{1}{2g_s}\big(\{X^i,X^a,X^b\}(\sigma^i)_{\dot{\alpha}}^{\ \
 \rho}(i\sigma^{ab})_{\beta}^{\ \ \theta}
 \psi_{\rho\theta}+\{X^i,X^j,X^a\}(\sigma^a)_{\beta}^{\ \
 \dot{\theta}}(i\sigma^{ij})_{\dot{\alpha}}^{\ \ \dot{\rho}}
 \psi_{\dot{\rho}\dot{\theta}}\big)\bigg]
\end{split}\ee%
where%
\be\label{tildeX}\begin{split}%
 \tilde{X}^{i}&=\mu
 p^+X^{i}+\frac{1}{3!g_s}\epsilon^{ijkl}\{X^{j},X^{k},X^{l}\}\cr
 \tilde{X}^{a}&=\mu
 p^+X^{a}+\frac{1}{3!g_s}\epsilon^{abcd}\{X^{b},X^{c},X^{d}\},
\end{split}\ee%
and similarly for the $Q_{\alpha\dot\beta}$.
The Hamiltonian is of course  given by \eqref{Hlc} and %
\bse\label{JijJab}\begin{align}%
 {\cal J}_{ij}&=\int d^3\sigma\biggl[X^iP^j-X^jP^i+\frac{1}{\mu
 p^+g^2_s}(P_E^iB^j-P_E^jB^i)-2\psi^{\dagger\alpha\beta}(i\sigma^{ij})^\rho_\alpha\psi_{\rho\beta}
 +2\psi^{\dagger\dot{\alpha}\dot{\beta}}(i\sigma^{ij})^{\dot{\rho}}_{\dot{\alpha}}\psi_{\dot{\rho}\dot{\beta}}\biggr]
 \\
 {\cal J}_{ab}& = \int d^3\sigma \biggl[X^aP^b-X^bP^a+\frac{1}{\mu
 p^+g^2_s}(P_E^aB^b-P_E^bB^a)-2\psi^{\dagger\alpha\beta}(i\sigma^{ab})^\rho_\beta\psi_{\alpha\rho}
 +2\psi^{\dagger\dot{\alpha}\dot{\beta}}(i\sigma^{ab})^{\dot{\rho}}_{\dot{\beta}}\psi_{\dot{\alpha}\dot{\rho}}
 \biggr]
\end{align}\ese%
and the extensions are obtained to be%
\bse\label{extensions}\begin{align}%
 {\cal
Z}&=\frac{1}{p^+}\int d^3\sigma \big[(P^I-i\tilde{X}^I)(\frac{B^I}{g_s}+iP_E^I)\big]\\%
 {\cal
Z}_{ijab}&=-\frac{i}{4p^+g_s}\int d^3\sigma
\big[(\frac{B^i}{g_s}+iP_E^i)\{X^j,X^a,X^b\}+ i,j \leftrightarrow\
a,b \big]\\
 {\cal R}_{ijab} &=\frac{\mu}{g_s}\int d^3\sigma \{X^i,X^j,X^a\}X^b
\end{align}\ese%
To verify the above commutations relations we have employed the
basic Poisson brackets:%
\bse\label{Poissonbrackets}\begin{align}%
[X^I(\sigma),P^J(\sigma')] &=i\delta^{IJ}\delta(\sigma-\sigma')\\
[A_r(\sigma), P_E^s(\sigma')] &=i\delta_r^s\delta(\sigma-\sigma')\\
\{\psi_{\alpha\beta}(\sigma),\psi^{\dagger\rho\lambda}(\sigma')\}
&=\delta_{\alpha}^{\rho}\delta_{\lambda}^{\beta}\delta(\sigma-\sigma')\\
\{\psi_{\alpha\beta}(\sigma),\psi^{\dagger}_{\rho\lambda}(\sigma')\}
&=\epsilon_{\alpha\rho}\epsilon_{\beta\lambda}\delta(\sigma-\sigma')\
,
\end{align}\ese
where (\ref{Poissonbrackets}b) is subject to $\partial_r P^r_E=0$.
We choose to fix the $U(1)$ symmetry in the  Coulomb gauge
$A_0=0$.

\subsection{1/2 BPS configurations}

From the superalgebra given in the previous section it is clear that
for the 1/2 BPS configurations of our model, those which preserve 16
(that is all of the) supercharges, the right-hand-side of the
fermionic anticommutators should vanish. That is, for 1/2 BPS
configurations we must have%
\be\label{half-BPS}%
{\cal H}=0, \quad {\cal J}_{ij}={\cal J}_{ab}=0, \qquad {\cal Z}=0,
\quad {\cal R}_{ijab}={\cal Z}_{ijab}=0.
\ee%
The above is only possible if we turn off fermions, $P^I=P^I_E=0$,
$B^I=0$ and $\tilde X^i=0,\ X^a=0$ or ${\tilde X}^a=0,\ X^i=0$
($\tilde X^i, \ \tilde X^a$ are defined in \eqref{tildeX}). These
two choices are related by the $i\leftrightarrow a$ exchange
$\mathbb{Z}_2$ symmetry, therefore we only consider the $\tilde
X^i=0,\ X^a=0$ case, that is \cite{Gianthedge}%
\be\label{spherical-solution}%
\{X^i,X^j,X^k\}=-\mu p^+g_s\ \epsilon^{ijkl}X_l\ .
\ee%

Eq.\eqref{spherical-solution} has two $SO(4)$ invariant solutions:\\
{\it i)} $X^i=0$ which specifies a zero size point like spherical
3-brane and\\
{\it ii)} A three sphere of radius $R$ \cite{Gianthedge}%
 \be\label{giant-radius}
 R^2=\mu p^+ g_s\ ,
 \ee
(if we recover $\alpha'$'s that is $R^2=2\pi \mu p^+\alpha' g_s$).
\footnote{It is worth noting that fixing the light-cone gauge from
the viewpoint of the above 1/2 BPS configurations corresponds to
going to the rest frame of these objects. {}From the background
plane-wave point of view, these are objects following the light-like
geodesic $\frac{\partial}{\partial X^-}$ with the momentum $p^+$
along the light-like trajectory. This also justifies the name giant
\emph{graviton}.}

To verify this it is enough to take%
\be\label{explicit-sphere}\begin{split}
 X^1=R\sin\psi\sin\theta\cos\phi &, \quad
 {X^2=R\sin\psi\sin\theta\sin\phi}\\
 {X^3=R\sin\psi\cos\theta} &, \quad {X^4=R\cos\psi}%
\end{split}\ee%
and recall that in the above coordinates
\[\{F, G, K\}=\frac{1}{\sin^2\psi\sin\theta} \epsilon^{rps}
\partial_r F\ \partial_p G\ \partial_s K\ ,\]%
where now $\epsilon^{rps}$ the totally antisymmetry Levi-Civita
tensor and takes only $0, \pm 1$ values. This spherical solutions
are the giant gravitons on the plane-wave background.

The fact that the above solutions are 1/2 BPS can also be seen
directly from the supercharges and that the supersymmetric
variations of fermions%
\bse\label{variation1}\begin{align} %
\delta\psi_{\rho\lambda}
&=i\{\epsilon^{\dagger\alpha\dot{\beta}}Q_{\alpha\dot{\beta}}+\epsilon_{\alpha\dot{\beta}}Q^{\dagger\alpha\dot{\beta}}
+\epsilon^{\dagger\dot{\alpha}\beta}Q_{\dot{\alpha}\beta}+\epsilon_{\dot{\alpha}\beta}Q^{\dagger\dot{\alpha}\beta}\
, \psi_{\rho\lambda}\} \\
\delta\psi_{\dot{\rho}\dot{\lambda}}&=i\{\epsilon^{\dagger\alpha\dot{\beta}}Q_{\alpha\dot{\beta}}+\epsilon_{\alpha\dot{\beta}}Q^{\dagger\alpha\dot{\beta}}
+\epsilon^{\dagger\dot{\alpha}\beta}Q_{\dot{\alpha}\beta}+\epsilon_{\dot{\alpha}\beta}Q^{\dagger\dot{\alpha}\beta}\
, \psi_{\dot{\rho}\dot{\lambda}}\}
\end{align}
\ese%
vanish for all sixteen possible supersymmetry transformation
parameters $\epsilon_{\alpha\dot{\beta}},\
\epsilon_{\dot\alpha{\beta}}$, once we plug in the above three
spherical solutions.
%

%
%
\section{BPS configurations with static electric fields}

In this section we study BPS configurations involving given
electromagnetic fields. The case of our interest is mainly the
static electromagnetic field, however,  we will also briefly
discuss the BPS electromagnetic waves. These configurations can be
classified by the amount of supersymmetry they preserve. A class
of less BPS configurations can be understood as deformations of
1/2 BPS states discussed in the previous section. For a general
(less) BPS state the supersymmetry transformations may vanish for
some specific choices for the supersymmetry transformation
parameters. The number of supersymmetries preserved is then the
number of independent \emph{real} $\epsilon_{\alpha\dot{\beta}},\
\epsilon_{\dot\alpha{\beta}}$'s which satisfy
$\delta\psi_{\rho\lambda}=0\,
,\delta\psi_{\dot{\rho}\dot{\lambda}}=0$ equation for that
specific configuration.

For example, consider the spherical 1/2 BPS giant graviton
configuration, but now turn on electric and magnetic fields, such that%
 \be\label{photon}%
 P^2_E=\frac{1}{g_s} B^1,
 \ee
and all the other components are zero. The above describes a 1/4
BPS state of a photon, for which ${\cal H}=\mu {\cal
J}_{12}=\frac{1}{p^+}P^2_E$ \cite{Gianthedge}. (Note that the time
dependence of the gauge field is given by its equations of motion
which for this case is basically the same as Maxwell equations on
$R\times S^3$ and we do not present them here explicitly.) One can
of course construct less BPS electromagnetic waves (photons) which
all propagate on the spherical three brane by superposing various
photon states propagating in different directions \cite{Kim}.

In the rest of this section we only focus on the static
electromagnetic fields. Requiring the configurations involving given
static electric or magnetic fields to be BPS, as we will see, forces
us to deform the shape of the three sphere.

Here we only study static configurations, that is we set $P^I=0$,
which are deformations of three sphere giants in the $X^i$
direction, that is we set $X^a=0$ and $P^a_E=B^a=0$, and of course
turn the fermions off. For this specific class the BPS condition
simplifies to
\bse\label{BPS-static}\begin{align}%
\delta\psi_{\rho\lambda}&=\big[i\tilde{X}^i\epsilon_{\dot{\alpha}\lambda}
-(\frac{B^i}{g_s}+iP^i_E)\epsilon^\dagger_{\dot{\alpha}\lambda}\big](\sigma^i)_{\
\ \rho}^{\dot{\alpha}}=0\\
\delta\psi_{\dot{\rho}\dot{\lambda}}&=\big[i\tilde{X}^i\epsilon_{\alpha\dot{\lambda}}
-(\frac{B^i}{g_s}+iP^i_E)\epsilon^\dagger_{\alpha\dot{\lambda}}\big](\sigma^i)_{\dot{\rho}}^{\
\ \alpha}=0\end{align}\ese%
For these configurations it is evident that ${\cal J}_{ab}=0$ and
${\cal R}_{ijab}={\cal Z}_{ijab}=0$. The only non-vanishing bosonic
generators can hence be ${\cal J}_{ij}$, ${\cal H}$ and ${\cal Z}$
({\emph{cf.} \eqref{JijJab} and \eqref{extensions}}). ${\cal H}$ is
positive definite and only vanishes for the three sphere giants and
for all these configurations ${\cal H}\neq 0$. As we will show, for
1/4 BPS configurations satisfying \eqref{BPS-static} ${\cal J}_{ij}$
also vanishes and for all of the 1/4 BPS configurations the BPS
condition is realized as ${\cal H}=\pm {\cal Z}$. In this sense we
consider new class of BPS solutions to the three brane giant
graviton theory which has not been studied in the literature before.
(In the literature mainly the configurations with non-vanishing
${\cal J}_{ij},\ {\cal J}_{ab},\ {\cal R}_{ijab}$ have been
considered e.g. see
\cite{Gianthedge, TGMT, tinysym, Das}).%

The BPS equations \eqref{BPS-static} are relating $\epsilon$\ and
$\epsilon ^\dagger$ and therefore they are only satisfied if
$\tilde X^i$ and $\Pi^i$,%
\be\label{Pi-i}%
\Pi^i\equiv \frac{B^i}{g_s}+i P^i_E,%
\ee%
are related in a specific way. For 1/4 BPS configurations this
happens if and only if
\begin{subequations}\label{1/4BPS-eqn}
\begin{align}
\Pi^i\bar\Pi^j&=\Pi^j\bar\Pi^i \\ %
\tilde X^i\tilde X^i &=\Pi^i\bar \Pi^i.
\end{align}
\end{subequations}
(\ref{1/4BPS-eqn}a) can also be written as%
\be\label{P||B}%
P^i_E B^j=P^j_E B^i%
\ee%
 and therefore for 1/4 BPS configurations ${\cal J}_{ij}=0$.
Note that once \eqref{1/4BPS-eqn} are fulfilled (\ref{BPS-static}a)
and (\ref{BPS-static}b) become identical. (A more detailed analysis
leading to the above equations may be found in Appendix
\ref{BPS-Appendix}.) Eq.\eqref{P||B} is satisfied if either $P^i_E$
or $B^i$ is vanishing or when $P_E$ is parallel to $B$ when both are
non-zero. In this paper we only consider the case with non-zero
$P^i_E$, $B^i=0$. The case with vanishing $P_E^i$ and non-zero $B$
can be obtained from the former using the electric-magnetic duality
\eqref{duality}.

For the pure electric case ($B^i=0$), \eqref{1/4BPS-eqn}
simplifies to%
\be\label{electric-BPS}%
\tilde X^i\tilde X^i= P^i_E P^i_E.%
\ee%
Since two $SO(4)$ vectors of the same norm are always related by
an $SO(4)$ rotation,%
\be\label{newbps1} %
\tilde{X}^i=R^{ij}P_E^j, \qquad R^{ij}R^{kj}=\delta^{ik}.%
\ee%
Before studying specific solutions, we also discuss the BPS
``perfect square trick''. For the cases with only non-zero $\tilde
X^i$ and
$P_E^i$  the Hamiltonian takes the simple form%
 \be\begin{split}%
 {\cal H}&=\frac{1}{2p^+}\int d^3\sigma \ [(\tilde
 X^i)^2+(P_E^i)^2]\cr
 &=\frac{1}{2p^+}\int d^3\sigma \
 [(\tilde{X}^i\pm R^{ij}P_E^j)^2\mp 2\tilde{X}^iR^{ij}P_E^j]
\end{split}\ee%
where $R_{ij}R_{kj}=\delta_{ik}$ is an $SO(4)$ rotation. The usual
BPS arguments then tells us that ${\cal H}$ is minimized when
$\tilde{X}^i=R^{ij}P_E^j$.

In the rest of this section we study solutions to \eqref{newbps1}
for given specific static electric fields. We analyze two class of
solutions. In section 3.1 we study  cases which are of the form of
giant three branes deformed as a result of the electric field. In
section 3.2 we study cases where we have string type configurations.
These configurations may be thought as extremely deformed three
branes which effectively behave like fundamental strings or
equivalently as deformations about $X=0$ vacuum.

%
%

\subsection{Giant-like configurations}
In this section we  turn on electric fields on the giant graviton
and study its shape deformation induced by the field. We consider
two cases, first the case where the electric field is sourced by
two equally charged but opposite point charges placed on the North
and South poles of the three sphere, and second we study the
constant electric field on the brane.

\subsubsection{BIGGons, BIons on the Giant Gravitons}
Consider the electric fields sourced by point charges on the three
sphere giants. Since the three sphere is compact we cannot place
non-zero net charge on it and hence the simplest possibility is an
electric dipole composed of a plus and a minus charge put on the
South and North poles of the three sphere giant. To make this
configuration BPS we need to turn on $X^i$ in a particular way,
dictated by the BPS equations \eqref{newbps1}. This generalizes the
BIons to the giant gravitons. This problem was first considered in
\cite{Gianthedge}, where it was only analyzed in the Hamiltonian
which is expanded up to quadratic order in the fields. Here we
intend to make the full Born-Infeld analysis, to all orders in
fields. Nonetheless as we will show the amount of supersymmetries
and bosonic symmetries of the system remains the same compared to
the second order analysis of the Hedgehog case.

Let us start by solving for the electric field:
\be%
 \vec{\nabla}.\vec{E}=\frac{1}{\sin^2\psi}\partial_\psi\big(\sin^2\psi
 E\big)=\frac{Q}{\sin^2\psi}\big(\delta(\psi)
 -\delta(\pi-\psi)\big)\delta(\cos\theta)\delta(\phi)%
\ee%
yielding%
\be\label{electric-dipole}%
 E^\psi=\frac{Q}{\sin^2\psi}\ .%
\ee%
The above electric field keeps the $SU(2)_D$ (which acts on
$\theta,\phi$ directions) and  $X^i$ should also be turned on
keeping the same $SU(2)_D$, {\it i.e.}
\be\begin{split}\label{X1-4}%
 X^1&=R S(\psi)\sin\theta\cos\phi\cr%
 X^2&=R S(\psi)\sin\theta\sin\phi\cr %
 X^3&=R S(\psi)\cos\theta\cr%
 X^4&=R C(\psi)%
\end{split}\ee %
The BPS equation \eqref{newbps1} is then written as
\be\label{CS1}\begin{split}%
 \sin^2\psi S+ S^2 C'=&\sigma\lambda(S'\cos\alpha+C'\sin\alpha)\cr%
 \sin^2\psi C- S^2 S'=&\sigma\lambda(C'\cos\alpha-S'\sin\alpha)
\end{split}\ee %
where we have used the definition of $P^i_E$, $P^i_E=E^\psi
\partial_\psi X^i$, $\alpha$ is the rotation angle relating
$\tilde X^i$ and $P^i_E$ and $\sigma$ is just a plus or minus
sign. (Although $\sigma$ could be absorbed in the definition of
rotation angle $\alpha$, it turns out to be more convenient to
keep $\sigma$.) Finally, $S'=\frac{dS}{d\psi},\ C'=
\frac{dC}{d\psi}$ and $\lambda=\frac{Q}{\mu}$. Without loss of
generality we take $\lambda$ to be positive.

It is readily seen that both side of above equations under the
parity transformation $\theta\ ,\ \psi\rightarrow \pi-\theta\ ,\
\pi-\psi$ and $\phi\rightarrow \pi+\phi$ behave in the same way if
under parity $\alpha\to \pi-\alpha$ and
 \[ S(\psi)=S(\pi-\psi),
\qquad C(\psi)=-C(\pi-\psi).\]

The BPS equations take a simpler form in terms of ``polar coordinate variable''
\be\begin{split}\label{CS}%
 S&=r(\psi)\sin\chi(\psi)\cr%
 C&=r(\psi)\cos\chi(\psi).%
\end{split}\ee %
For $Q=0$ case it is evident that $r(\psi)=1$ and $\chi=\psi$ is a
solution to \eqref{CS1}. Deviation of $\chi$ from $\psi$ then
comes from the charges we have in the system. Under parity $\chi$
should also behave the same as $\psi$, {\it i.e.}
\[\chi(\psi)=\pi-\chi(\pi-\psi)\]
and $r(\psi)=r(\pi-\psi)$.

In the BPS equations $\alpha$ is an arbitrary angle but should
transform suitably under parity. With the choice
\[
\alpha=\chi
\]
the BPS equations, which are non-linear coupled first order
differential equations for $r$ and $\chi$ take a simple form and
could be solved. With this choice straightforward algebra leads to
\be\label{r-S}%
 r=1-\frac{\sigma\lambda}{S}
\ee%
where we have used the initial condition that for $\lambda=0$,
$r=1$. Using the above one can eliminate $r$ to obtain the
equation for $S$ (or $\chi$)
\be\label{x-psi}%
\sigma' \sin^2\psi=\bigg(\frac{S^5-3\sigma\lambda S^4+\lambda^2
S-\sigma\lambda^3}
 {S(S-\sigma\lambda)\sqrt{S^2+\lambda^2-2\sigma \lambda S-S^4}}\bigg) S'%
\ee%
where $\sigma'$ is +1 (-1) for $\psi<\pi/2$ ($\psi>\pi/2$).
Although the shape of the giant is completely determined by
\eqref{r-S}, equation \eqref{x-psi} is needed to find the range
over which the variable $S$ is varying. Recalling that
$\sin\chi=r/S$ from \eqref{r-S} we learn that%
 \be\label{S-range}%
 0\leq\sin\chi=\frac{S^2}{S-\sigma\lambda}\leq 1
\ee%
and hence%
\be\label{s}%
 \frac{1}{2}(1-\sqrt{1-4\sigma\lambda})\leq S \leq\frac{1}{2}(1+\sqrt{1-4\sigma\lambda}).%
\ee%

Let us now consider the $\sigma=+$ and $\sigma=-$ cases
separately:\\
$\bullet\ \ \ \sigma=-$, {\it the outward spike}:

For this case \eqref{r-S} and \eqref{s} read as%
\be\label{sigma-1}%
r=1+\frac{\lambda}{S}, \qquad  0\leq S \leq\frac{1}{2}(1+\sqrt{1+4\lambda}),%
\ee%
and hence $r>1$. If we ignore  \eqref{x-psi}, \eqref{sigma-1}
describes a three sphere with two spikes {\it going off to
infinity} coming out of the two poles. The lowest value $S$ can
take $S_0$, is however, further restricted by \eqref{x-psi} and
$S$ can never become zero. Hence,the spikes are cut off and do not
go to infinity. This is in fact the main qualitative difference of
the full Born-Infeld analysis compared to the case of
\cite{Gianthedge}.

To see this let us suppose that $S$ can become arbitrarily small
and study \eqref{x-psi} in the $S\to 0$ region which while keeping
$\lambda$ fixed, necessarily happens when $\psi\to 0$. In this
limit equation \eqref{x-psi} implies that
\be%
 S\sim k e^{-\frac{1}{3!\lambda}\psi^3},
\ee%
where $k$ is an integration constant. This is a contradiction, as
for small $\psi$ and finite $k$,  $S$ does not approach zero.
Therefore, $S$ cannot become smaller than some $S_0$ where the
spike is cut off. $S_0$ is a complicated function of $\lambda$
which in principle can be computed integrating \eqref{x-psi}.
However, from \eqref{x-psi}, one can deduce that when $\lambda\to
0$, $S_0(\lambda)$ also tends to zero.

 In sum, the shape of the brane is given by
\be\label{sigma-2}%
r=1+\frac{\lambda}{S}, \qquad  S_0(\lambda) \leq S \leq\frac{1}{2}(1+\sqrt{1+4\lambda}).%
\ee%
This function is depicted in Figure 1.\\
$\bullet\ \ \ \sigma=+$, {\it the inward spike}:

For this case \eqref{r-S} and \eqref{s} read as%
\be\label{sigma+1}%
r=1-\frac{\lambda}{S}, \qquad
\frac{1}{2}(1-\sqrt{1-4\lambda})\leq S \leq
\frac{1}{2}(1+\sqrt{1-4\lambda}),%
\ee%
and hence always $r<1$. Since $r$ cannot take negative values
$S\geq \lambda$, furthermore in order \eqref{S-range} to hold we
must have $\lambda \leq 1/4$. For $\lambda>1/4$ there is no BPS
inward spike solution. If $\lambda \leq 1/4$ then
$\frac{1}{2}(1-\sqrt{1-4\lambda}) \leq \lambda$ and hence $S$ can
take all the values in the range given in \eqref{sigma+1}. For the
maximal and minimal values of $S$, it is easily seen that
$r_{min}=S_{min},\ r_{max}=S_{max}$ and that%
\[\frac{1}{2}(1-\sqrt{1-4\lambda})\leq r,\ S \leq
\frac{1}{2}(1+\sqrt{1-4\lambda}).\] %
Therefore, the size of the throat in the spike is $S_{min}$.
Moreover, at $S_{min}$ and $S_{max}$  the slope of the curve is
infinite ({\it i.e.} the tangent is parallel to vertical axes). The
inward spike has the topology of $S^2\times S^1$ and has been
depicted in Figure 2.

For $\lambda=1/4$, the only value that $S$ or $r$ can take is
$r=S=1/2$ and the giant becomes a two-sphere of radius $1/2$. If
$\lambda$ is larger than $1/4$  the solution is not smoothly
connected to the spherical three brane giant.

\begin{figure}[ht]
\includegraphics[scale=1]{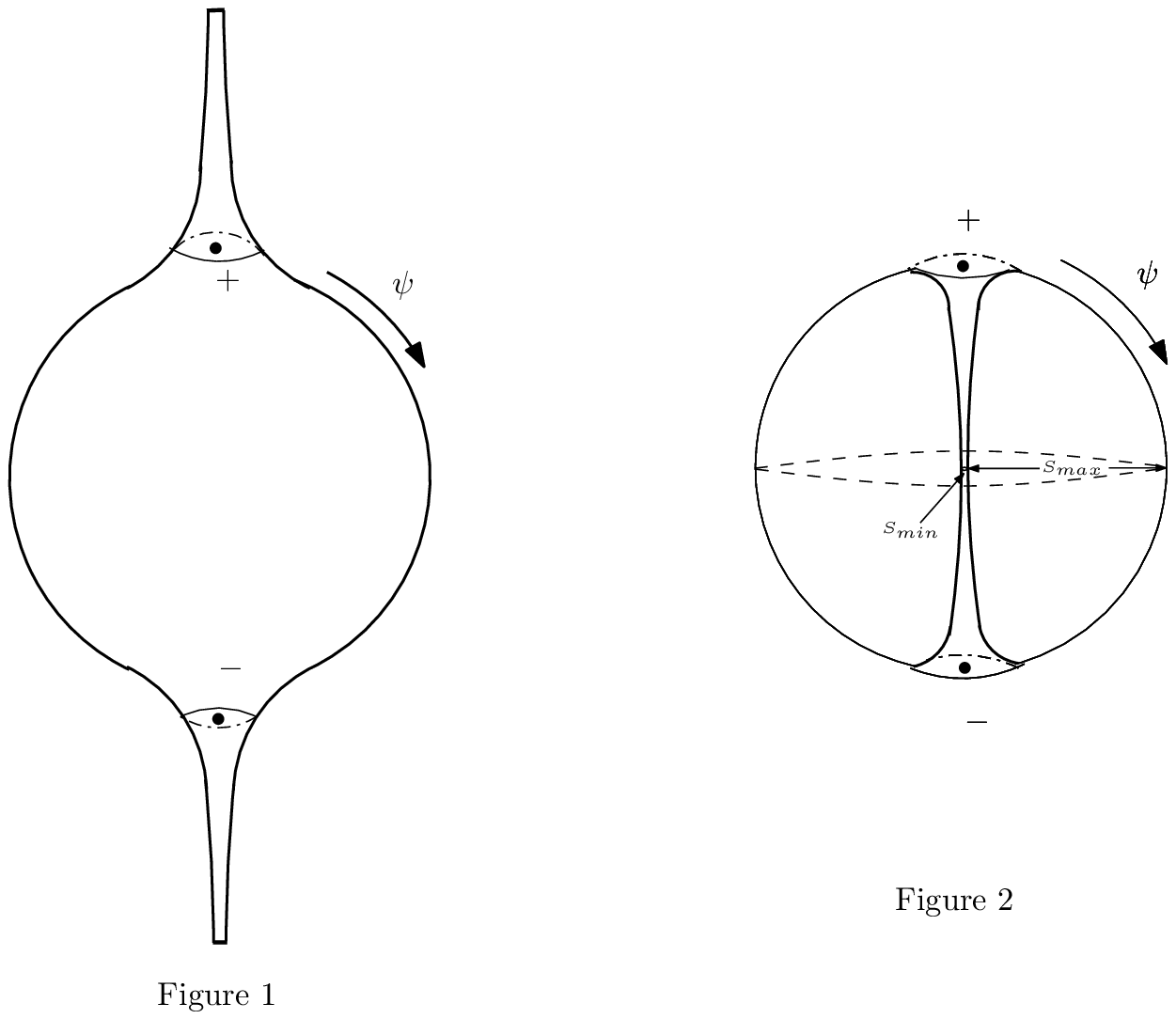}%
\caption{The shape of the outward BIGGons which is obtained for the
choice $\sigma=-$ in \eqref{r-S}. The spike is cut off and does not
go to infinity. The minimum value of $r$ is obtained when $S$ takes
its maximal value,
$r_{min}=S_{max}=\frac{1}{2}(1+\sqrt{1+4\lambda})$. The maximum
value of $r$, which is basically (one plus) the height of the spike
is $1+\lambda/S_0$. Note that at $S=S_0$ the slope of the spike is
large but still finite.
\newline Figure 2: The $\sigma=+$ in \eqref{r-S} gives the inward BIGGon.
This case is obtained only if $\lambda\leq 1/4$. The minimum and
maximum value of the radius $r$ is given by the minimum and
maximum values of $S$, respectively,
$\frac{1}{2}(1-\sqrt{1-4\lambda})$ and $
\frac{1}{2}(1+\sqrt{1-4\lambda})$. }
\end{figure}%

It is instructive to compare our analysis to the case of giant
Hedgehog \cite{Gianthedge} which is obtained from our equations if
we first take $\lambda\ll 1$ limit. In this limit one can drop the
non-linear terms in \eqref{CS1} and  $\chi=\alpha\simeq \psi$. In
this case the spikes go off to infinity.  For generic values of
$\psi$ (when $\psi$ is not close to $0$ or $\pi$) the value of the
electric field \eqref{electric-dipole} as well as $\partial_\psi
X^i$ (or $S'$ and $C'$) are small and the quadratic approximation
in the Hamiltonian is a good one. Close to $\psi=0$ or $\psi=\pi$
region, however, we start seeing deviations of Born-Infeld from
the quadratic analysis and the Maxwell approximation is not a
valid one. As discussed the non-linear terms around $\psi=0$
become dominant and cap off the spike. This is in contrast to the
flat brane case where for both of the second order Maxwell theory
and the full Born-Infeld cases the spike is infinite
\cite{Callan-Maldacena, Gibbons}.

As a result of the deformation in the shape of the three sphere
giant the singularity of the point charges has been removed. To
see this let us work out the value of the norm of the electric
field:%
\be\label{electric-pullback}%
\begin{split}%
E^2&=(E^\psi)^2\cdot g_{\psi\psi}=\frac{Q^2
R^2}{\sin^4\psi}(C'^2+S'^2)\cr%
 &= Q^2 R^2\frac{(S-\sigma\lambda)^3}{S^4(S^5-3\sigma\lambda
 S^4+\lambda^2 S-\sigma\lambda^3)}.
 \end{split}
 \ee%
The factor in the denominator does not vanish for neither of the
inward and outward spikes (note that $S$ never becomes zero). The
energy density of the solution,
\be%
 {\cal{H}}=p^+\int  d\psi d\theta d\phi\ \sin^2\psi\sin\theta\ E^2,
\ee%
where $E^2$ is the expression given in \eqref{electric-pullback}, is
hence finite.

%
%
\subsubsection{Squashed Giant configuration}

As the second example let us consider the case with a given
constant electric field on the brane%
 \be
 P_E^i=P_E^r\partial_rX^i=P_E^\phi\partial_\phi X^i,\ \ P_E^\phi=p^+E=constant
\ee%
with this electric field the BPS equation \eqref{newbps1} takes
the form
\be\label{BPS-eq-squashed}\begin{split}%
 \tilde{X}^1&=\sigma(P^1_E\cos\alpha+P^2_E\sin\alpha)\cr%
 \tilde{X}^2&=\sigma(P^2_E\cos\alpha-P^1_E\sin\alpha)\cr%
 \tilde{X}^3&=0\cr%
 \tilde{X}^4&=0%
\end{split}\ee%
where again $\sigma$ takes $\pm 1$ values and $\alpha$ is a
yet-to-be-fixed angle. In the following we choose $E$ to be
positive. Equation \eqref{BPS-eq-squashed} is solved with
\be\begin{split}%
 X^1&=Ra\sin\psi\sin\theta\cos\phi\cr%
 X^2&=Ra\sin\psi\sin\theta\sin\phi\cr%
 X^3&=Rb\sin\psi\cos\theta\cr%
 X^4&=Rb\cos\psi%
\end{split}\ee%
provided that%
\be\label{a-b-squashed}
 b^2=1-\sigma\frac{E}{\mu}\ \ ,\ \ a^2=1\ \ ,\ \ \alpha=\pi/2.
\ee%

The above solution describes a brane with a deformed
sphere ellipsoidal shape. Shape of the three brane can be easily worked out%
\be
 X_1^2+X_2^2+\frac{1}{1-\sigma E/\mu}(X^2_3+X^2_4)=R^2.
\ee%
For $\sigma=-1$ the configuration exist for all values of electric
field. For $\sigma=+1$ only for $E<\mu$ we have an ellipsoid, for
$E=\mu$ the shape is singular (see section 3.2 for more details)
and for $E>\mu$ the brane has a hyperboloid shape. For both of the
signs, when we have ellipsoids, there is a 2-sphere cross section.
To see this set $X_3=\sqrt{1-\sigma E/\mu}\ r\sin\gamma$ and
$X_4=\sqrt{1-\sigma E/\mu}\ r\cos\gamma$. Then we recover a
2-sphere of radius $R$ in $r12$-space. This shows the $SU(2)$
isometry of the solution. There is a circular cross section e.g.
at $X^3=X^4=0$ with radius $R$. Therefore our $1/4$ BPS
configuration keeps $SU(2)\times U(1)$ isometries out of the whole
$SO(4)$. Similar ellipsoid branes can also been obtained from a
rotating three sphere giants \cite{tinysym}.
Total energy for these configuration can be evaluated%
\be\label{e-squashed}
 {\cal{H}}=\int d^3\sigma p^+E^2\big((\partial_\phi X^1)^2+(\partial_\phi
 X^2)^2\big)= \pi^2 \mu g_s  (p^+ E)^2.
\ee%

%
%
\subsection{String-like configurations}
In this section we study 1/4 BPS string-like solutions of
\eqref{Hlc} which involve constant background electric field. As
discussed \eqref{Hlc} has two kind of 1/2 BPS configurations, a
single giant three sphere of radius $R^2=\mu p^+ g_s$, or a zero
size brane siting at $X^i=X^a=0$. The string-like solutions of this
section can then be understood either as extremely deformed
(squashed) three branes or as deformation about the $X=0$ solution.
Here we consider examples of each kind in the background electric
field
\be\label{electric}%
 P^i_E=P_E^r\partial_r X^i=P_E^\phi\partial_\phi X^i\  ,\
 P_E^\phi=p^+E=constant
\ee%
while $X^3=X^4=0$.

For this choice, the BPS equation \eqref{newbps1} becomes%
\be\label{string-BPS}\begin{split}%
 R^2X^1=\sigma g_s(P^1_E\cos\alpha+P^2_E\sin\alpha)\cr
 R^2X^2=\sigma g_s(P^2_E\cos\alpha-P^1_E\sin\alpha)
\end{split}\ee%
%
%
\subsubsection{Circular string}
In order to find other string-like BPS solution we start with%
\be\label{circular-string}%
 X^1=Ra\cos\phi\ ,\
 X^2=Rb\sin\phi%
\ee%
(while $\psi=\theta=\frac{\pi}{2}$). The above solves
\eqref{string-BPS} if \be
 E^2=\mu^2\ \ ,\ \ a^2=b^2\ \ ,\ \ \alpha=\pi/2.
\ee%
This configuration which is a circular closed string of radius $Ra$,
is a special case of the squashed giant of section 3.2.1 with
$\sigma=+$ and $E=\mu$. In this sense this solution is an example of
extremely deformed three brane giant. In another viewpoint if we
turn off the electric field this solution reduces to the $X=0$
vacuum. The total energy density of this configuration is ${\cal
E}=\mu g_s a^2 (p^+ E)^2$.

This configuration is in fact a fundamental string in the
$X^1,X^2$ plane, in the pp-wave
background. To see it we consider fluctuations around this solution%
\be\label{fluct-string}%
 X^1=X_0^1+Y^1\ ,\
 X^2=X_0^2+Y^2
\ee%
where $X_0^i$ are the the circular string solution given in
\eqref{circular-string}. The Hamiltonian for these fluctuations is
\be\label{fluct-circular-string}\begin{split}%
 H=&\frac{1}{2p^+}(P_1^2+P_2^2+P_3^2+P_4^2)+\frac{1}{2}\mu^2 p^+(X_3^2+X_4^2)\cr%
 +&\frac{1}{2}\mu^2 p^+(Y_1^2+Y_2^2)+\frac{\mu^2p^+}{2}\big((\partial_\phi Y_1)^2+(\partial_\phi Y_2)^2\big).
\end{split}\ee%
which is the light-cone Hamiltonian for a four dimensional string in
the plane-wave background. The tension term, the last term, is
coming from the electric field. The tension of this string is
$\mu=E$. This string can be thought as an array of tiny electric
dipoles which are aligned in the background electric field. In this
picture it becomes clear that the tension is proportional to the
electric field and that string is only in the plane where the
electric field is turned on.

%
%
\subsubsection{Stretched string}

\eqref{circular-string} solves \eqref{string-BPS} with another
value for $\alpha$:%
\be\label{straight-string}
 a(\phi)\cos\phi=ce^{\pm\frac{\mu}{E}\phi}\ ,\
 b(\phi)\sin\phi=de^{\pm\frac{\mu}{E}\phi}\ ,\quad \alpha=0
\ee%
with no restriction on $E$. Physically for vanishing electric
field we should recover the $X=0$ vacuum, therefore {plus} sign in
the above solution is not acceptable. The above describes a
straight string along the
\be%
 X^2= \frac{d}{c} X^1
\ee%
line with $ Rce^{-\frac{2\pi \mu}{E}}\leq X^1\leq Rc$.
The string length is equal to%
\be%
 l=R(c^2+d^2)^\half (1-e^{-\frac{2\pi\mu}{E}})^{\half}.%
\ee%
The maximum possible length is then obtained for $E\to \infty$.
The energy density of this configuration is
${\cal{E}}=\mu^2p^+l^2$.

This configuration is describing a fundamental string of tension
$E$. To see this one can work out the Hamiltonian for the
fluctuation about this solution. Inserting \eqref{fluct-string} into
the Hamiltonian we obtain an expression similarly to
\eqref{fluct-circular-string} but now the tension is equal to the
electric field $E$. This kind of string, similarly to the previous
case,  is composed of a set of electric dipoles which are ordered in
opposite direction to the electric field  $E$.
%
%
\subsection{Superalgebra viewpoint}\label{superalgebra}

To complete our BPS analysis we also study the BPS configurations
we have discussed directly from the superalgebra point of view. As
we discussed in general the rotation angle $\alpha$ is not a
constant and can in general be a function of $\psi,\ \theta$ or
$\phi$. To include the effects of this angle into the
superalgebra, therefore we need to modify the supercharge
densities given in \eqref{Q} to include the rotation $R_{ij}$. It
is straightforward to check that anti-commutator of the
supercharges
 \be\label{hatQ}\begin{split}
 \hat{Q}_{\dot{\alpha}\beta}=\frac{1}{\sqrt{2p^+}}\int d^3\sigma
 \bigg[&(P^i-i\tilde{X}^i)(\sigma^i)_{\dot{\alpha}}^{\ \ \rho}
 \psi_{\rho\beta}+(\frac{B^i}{g_s}+iP_E^i)R^{ij}(\sigma^j)^{\rho}_{\
 \ \dot{\alpha}}\psi^{\dagger}_{\rho\beta}\cr
 +&(P^a-i\tilde{X}^a)(\sigma^i)_{\beta}^{\ \ \dot{\rho}}
 \psi_{\dot{\alpha}\dot{\rho}}+(\frac{B^a}{g_s}+iP_E^a)R^{ab}(\sigma^b)^{\dot{\rho}}_{\
 \ \beta}\psi^{\dagger}_{\dot{\alpha}\dot{\rho}}\cr
 -&\frac{1}{2g_s}\big(\{X^i,X^a,X^b\}(\sigma^i)_{\dot{\alpha}}^{\ \
 \rho}(i\sigma^{ab})_{\beta}^{\ \ \theta}
 \psi_{\rho\theta}+\{X^i,X^j,X^a\}(\sigma^a)_{\beta}^{\ \
 \dot{\theta}}(i\sigma^{ij})_{\dot{\alpha}}^{\ \ \dot{\rho}}
 \psi_{\dot{\rho}\dot{\theta}}\big)\bigg]
\end{split}\ee%
where $R_{ij}$ is defined in \eqref{newbps1}, with its complex
conjugate reproduces the light-cone Hamiltonian but with modified
${\cal J}_{ij},\ {\cal J}_{ab},\ {\cal R}_{ijab}$. One can also
work out anti-commutators of two supercharges to read the central
extension $\hat{{\cal Z}}$. For the 1/4 BPS configuration of our
interest where only ${\cal H}$, $\hat{{\cal Z}}$ are
non-vanishing, %
\be\label{susyqqdagger2}
 \{\hat{Q}_{\dot{\alpha}\beta},\hat{Q}^{\dagger\ {\dot{\rho}\lambda}}\}=
{\cal H}\delta_{\dot\alpha}^{\dot\rho}\delta_\beta^\lambda\
,\qquad \{\hat{Q}_{\alpha\dot{\beta}},\hat{Q}^{\dagger\
\rho\dot{\lambda}}\}={\cal H}
\delta_{\alpha}^{\rho}\delta_{\dot{\beta}}^{\dot{\lambda}}
\ee%
and \be\label{susyqq2}
 \{\hat{Q}_{\dot{\alpha}\beta},\hat{Q}_{\dot{\rho}\lambda}\}= {\cal
 \hat{Z}}\epsilon_{\dot{\alpha}\dot{\rho}}\epsilon_{\beta\lambda}\
 ,\qquad
 \{\hat{Q}_{\alpha\dot{\beta}},\hat{Q}_{\rho\dot{\lambda}}\}={\cal
 \hat{Z}}\epsilon_{\alpha\rho}\epsilon_{\dot{\beta}\dot{\lambda}}
\ee%
where%
\be\label{extensions2}%
 {\cal\hat{Z}}=\frac{1}{p^+}\int d^3\sigma \big[(P^i-i\tilde{X}^i)R^{ij}(\frac{B^j}{g_s}+iP_E^j)+(P^a-i\tilde{X}^a)R^{ab}(\frac{B^b}{g_s}+iP_E^b)\big].%
\ee%
${\cal\hat{Z}}$ as well as the Hamiltonian ${\cal H}$ take different
values for each configurations. However, it is readily seen that for
all the configurations studied in this section ${\cal H}=\pm
{\cal{\hat{Z}}}$. To see the BPS nature of these configurations
explicitly, we note that the superalgebra produced by the
supercharges%
\bse\label{Qtilde} \begin{align}
\tilde{Q}_{\dot{\alpha}\beta}&=\frac{1}{\sqrt{2}}\big(\hat{Q}_{\dot{\alpha}\beta}\pm\epsilon_{\dot{\alpha}\dot{\rho}}
 \epsilon_{\beta\rho}\hat{Q}^{\dagger\dot{\rho}\rho}\big)
\\
\tilde{Q}_{\alpha\dot{\beta}}&=\frac{1}{\sqrt{2}}\big(\hat{Q}_{\alpha\dot{\beta}}\pm\epsilon_{\alpha\rho}
 \epsilon_{\dot{\beta}\dot{\rho}}\hat{Q}^{\dagger\rho\dot{\rho}}\big)
\end{align}\ese
is of the form
\bse\begin{align}%
 \{\tilde{Q}_{\dot{\alpha}\beta},\tilde{Q}^{\dagger\dot{\rho}\lambda}\}=
 \delta_{\dot{\alpha}}^{\ \ \dot{\rho}}\delta_\beta^{\ \ \lambda}({\cal{H}}\pm \cal{\hat{Z}})\quad &, \quad%
 \{\tilde{Q}_{\alpha\dot{\beta}},\tilde{Q}^{\dagger\rho\dot{\lambda}}\}=\delta^{\
 \ \rho}_\alpha\delta^{\ \ \dot{\lambda}}_{\dot{\beta}}
 ({\cal{H}}\pm \cal{\hat{Z}})\\
 \{\tilde{Q}_{\dot{\alpha}\beta},\tilde{Q}_{\dot{\rho}\lambda}\}= 0
 \quad &, \quad
 \{\tilde{Q}_{\alpha\dot{\beta}},\tilde{Q}_{\rho\dot{\lambda}}\}=0.
\end{align}\ese%
(Note that each of the plus and minus signs in \eqref{Qtilde} is
giving four independent supercharges and hence we need to consider
both signs to capture the whole superalgebra.) Therefore the
right-hand-side of the above superalgebra vanishes for eight (half
of) supercharges if ${\cal H}=\cal{\hat{Z}}$.

%
%

\section{Discussion}
In a quest to enhance our understanding of dynamics of D3-branes in
the plane-wave background we have constructed 1/4 BPS configurations
of such D-brane using the light-cone Hamiltonian and the
corresponding (dynamical) superalgebra. The BPS configurations we
studied in this work all involve static electric field of the brane.

We first studied the BPS configuration corresponding to the
electric field sourced by an electric dipole on the three sphere
giant graviton, generalizing BIons to spherical branes. The
electric dipole is composed of two opposite point charges $Q$
placed on the North and South poles of the brane. In contrast to
the flat brane BIons, due to non-linear Born-Infeld dynamics, the
(double) spikes are capped  off and do not extend to infinity. The
details of the configuration, e.g. the length of the spikes or the
size of their throats, are controlled by parameter $\lambda=Q/\mu$
($\mu$ is the scale associated with the background plane-wave). We
also discussed that besides the outward double spike
configuration, we can also have inward spikes, spikes piercing
through the three sphere giant (see the Figure). The inward spike
solution, however, only exists for $\lambda\leq 1/4$. Although we
start with placing two point charges on the three sphere giant,
and hence the corresponding electric field is singular, the shape
of the brane is deformed in a specific manner such that the final
configuration in the end is smooth with no singularity. This is
important because in principle the DBI action is not capturing all
the $\alpha'$-corrections to the brane dynamics and there are
higher order derivative $\alpha'$-corrections to DBI
\cite{Wyllard}. For our solutions, however, higher derivative
terms are subleading to the same $\alpha'$ order already present
in the DBI.

It is of obvious interest to study and analyze similar double spike
solutions from the dual ${\cal N}=4$ SYM theory as well as the tiny
graviton matrix theory \cite{TGMT, tinysym}. For the former, it is
notable that the ``Fat Magnons'' of \cite{shinji}, which are bound
states of ``giant magnons'' of \cite{Maldacena-Hofman} and giant
gravitons, are indeed the same object as our double spike solutions.
 The fact that in the full Born-Infeld
description the spikes turn out to have finite length and energy is
compatible with the expectation of constructing open string
excitations of the giants using the BMN type construction discussed
in \cite{Electric-Giants}. We hope to elaborate further on this
question in upcoming publications.

Besides the double spike solutions we also studied squashed giant
configurations and as showed there is a critical value for the
electric field, $E=\mu$, where the shape of the giant becomes
singular. It is desirable to understand better this critical
electric field both from the brane theory and the dual ${\cal N}=4$
SYM theory viewpoint.

We have also analyzed another class of 1/4 BPS configurations, the
stringy solutions. As discussed these strings which are in fact
deformations about the $X=0$ vacuum of the theory, can be understood
as follows. Each string is made out of (infinite number of) tiny
electric dipoles, which is the absence of an external electric
field, which due to the harmonic oscillator potential well provided
by the background plane-wave,  are all sitting on top of each other
at $X=0$. When the electric field is turned on, these dipoles are
all aligned in the direction of the electric field and hence the
tension of these strings are proportional to the electric field. In
the directions orthogonal to the electric field, because of the
harmonic oscillator potential coming from the background plane-wave
metric, we are still dealing with point like objects. This picture
is a very interesting one, suggesting that the strings on the
background plane-wave are  made out of ``string bits'' and the
string bits in their own turn are tiny, dipole like three spheres
and a fundamental string is in fact an (electric) flux tube. This
dipole interpretation gives a realization of the string bit
intuition coming from the BMN analysis. Moreover, this is also
compatible with the picture developed in the tiny graviton matrix
theory (TGMT) \cite{TGMT}, according which the tiny three sphere
branes, tiny gravitons, each carrying one unit of the light-cone
momentum are the building blocks of the type IIB strings on the
$AdS_5\times S^5$ or the plane-wave backgrounds. The other
interesting question regarding the string-like 1/4 BPS
configurations discussed here is to compare them with the ``giant
magnon''  \cite{Maldacena-Hofman} or the ``fat magonon''
\cite{shinji} configurations, though after taking the Penrose limit.

In this work we only discussed 1/4 BPS configurations involving only
electric field. As we showed, we have similar solutions in which the
electric field is replaced by magnetic field (according to
\eqref{duality}). It is also possible to have 1/4 BPS configurations
involving both static electric and magnetic  fields, which we did
not analyze here. Besides the 1/4 BPS solutions, we can have more
less BPS (1/8 BPS) configurations with more general electric and
magnetic fields. Analysis of these configurations is postponed to
future works.

%
%
\section*{Appendices}
\appendix
\section{Polyakov form of the $D_p$-brane action}\label{Polyakov}
Dirac-Born-Infeld (DBI) action which describes dynamics of a
$D_p$-brane can be put in another useful form by introducing an
additional field on the world-volume, an independent world-volume
metric. This form is useful in fixing the light-cone gauge (\emph{cf}. footnote 2). DBI-action is presented by%
\be
 S=-\frac{T}{g_{s}}\int d^{p+1}\sigma\sqrt{-\det\big(g_{\hat{\mu}\hat{\nu}}+F_{\hat{\mu}\hat{\nu}})}
\ee%
where $\hat{\mu}=0,..,p$ . Recalling the symmetry and antisymmetry
of $g_{\hat{\mu}\hat{\nu}}$ and $F_{\hat{\mu}\hat{\nu}}$, we can write %
\be\label{det-identity}
 \big[\det(g_{\hat{\mu}\hat{\nu}}+F_{\hat{\mu}\hat{\nu}})\big]^{\frac{1}{2}}= \big[\det(g_{\hat{\mu}\hat{\nu}}-F_{\hat{\mu}\hat{\nu}})\big]^{\frac{1}{2}}=
 \big[\det(g_{\hat{\mu}\hat{\nu}})\det(g_{\hat{\mu}\hat{\nu}}-F_{\hat{\mu}\hat{\alpha}}g^{\hat{\alpha}\hat{\beta}}
 F_{\hat{\beta}\hat{\nu}})\big]^\frac{1}{4}
\ee%

Next consider the action \cite{0606079}
\be\label{polyakov-action}%
 S^{\prime}=-\frac{T}{g_s}^{\prime}\int{d^{p+1}\sigma}\det(h
 g)^\frac{1}{4}\big[h^{\hat{\mu}\hat{\nu}}(g-F^2)_{\hat{\mu}\hat{\nu}}
 +\Lambda\big]
\ee%
where $h$ is a dynamical worldvolume metric and
$F^2_{\hat{\mu}\hat{\nu}}=F_{\hat{\mu}\hat{\alpha}}g^{\hat{\alpha}\hat{\beta}}
F_{\hat{\beta}\hat{\nu}}$.  Upon eliminating the $h$ field using
its equation of motion, $S^\prime$ reduces to the DBI action. To
see this we note that the equation of motion of the $h$ field is
obtained by setting the energy-momentum tensor, which  by
definition is the
variation of the action with respect to the worldvolume metric,  equal to zero. That is,%
\be\label{e-m}%
 T_{\hat{\mu}\hat{\nu}}=\frac{\delta S^{\prime}}{\delta
h^{\hat{\mu}\hat{\nu}}}=(g-F^2)_{\hat{\mu}\hat{\nu}}-\frac{1}{4}h_{\hat{\mu}\hat{\nu}}
 \big[h^{{\hat\alpha}\hat{\beta}}(g-F^2)_{\hat{\alpha}\hat{\beta}}
 +\Lambda\big]=0
\ee%
$\Lambda$ can be identified taking the trace of the above equation:%
 \be\label{Lambda}
 \Lambda=\frac{3-p}{p+1}h^{\hat{\mu}\hat{\nu}}(g-F^2)_{\hat{\mu}\hat{\nu}}.%
\ee
\eqref{Lambda} and \eqref{e-m} then yield%
\be\label{det}%
 \det\big((g-F^2)_{\hat{\mu}\hat{\nu}}\big)=
 \det(h_{\hat{\mu}\hat{\nu}}) \biggl[\frac{h^{{\hat\alpha}\hat{\beta}}(g-F^2)_{\hat{\alpha}\hat{\beta}}}{p+1}\biggr]^{p+1}.%
\ee%
Using the above and \eqref{det-identity} we recover the DBI-action
once we insert \eqref{det} into Polyakov action
\eqref{polyakov-action} (note that as is seen from \eqref{Lambda}
$h^{\hat{\mu}\hat{\nu}}(g-F^2)_{\hat{\mu}\hat{\nu}}$ is a constant
and not a variable or field.) It is also notable that for the case
of our interest, {\it i.e.} $p=3$, $\Lambda$ vanishes.
%
%
\section{Derivation of the light-cone Hamiltonian in more detail}\label{light-cone-Appendix}
To obtain the light-cone Hamiltonian we note that due to the local
diffeomorphism invariance the Hamiltonian ${\bf {\tilde {\cal H}}}$%
\be\label{tilde-H}%
{\bf {\tilde {\cal H}}}=\sum_\alpha \frac{\partial
L}{\partial\dot\Phi_\alpha}\dot\Phi_\alpha-L
\ee%
with $\Phi_\alpha\in\{ X^+, X^-, X^I,\ A_0, A_r, \psi,\bar\psi\}$,
should vanish for all physical configurations. Besides ${\bf {\tilde
{\cal H}}}=0$, in the light-cone gauge one should also impose
\eqref{Level-match} on the physical configurations. In the
light-cone gauge, after imposing \eqref{lightcone-time} and
\eqref{kappa-sym} we have  \cite{Kamimura}%
\be\label{tilde-H=0}\begin{split}%
{\bf {\tilde {\cal H}}}&= p^+\partial_\tau X^-+ P_I\dot
X^I+P^-+\frac{\partial L}{\partial F_{0r}} F_{0r}+\frac{\partial
L}{\partial \dot\psi}\dot\psi+\frac{\partial L}{\partial
\dot{\bar\psi}}\dot{\bar\psi}-L\\
&=0
\end{split}
\ee%
where we have also imposed the $U(1)$ gauge theory constraint%
\be\label{prE}
\partial_r P^r_E=0\ ,
\ee%
and%
\be\label{PI-def}
 P^I=\frac{\partial{\cal L}}{\partial \dot{X}_I}=-p^+\dot{X}^I
\ee%
\be\label{PrE-def}
 P^r_E=\frac{\partial{\cal L}}{\partial
 F_{0r}}=\frac{1}{g_s}\sqrt{-detN}N^{0r}.
\ee%
In the above two equations \eqref{Level-match}, that is
$N^{0r}=-N^{r0}$, has been used.
{}From \eqref{tilde-H=0} we learn that%
 \be\label{H-lc-P-}
H_{lc}\equiv P^-= -\biggl[p^+\partial_\tau X^-+ P_I\dot X^I+P^r_E
F_{0r}+\frac{\partial L}{\partial \dot\psi}\dot\psi+\frac{\partial
L}{\partial \dot{\bar\psi}}\dot{\bar\psi}-L\biggr]%
\ee%

We should now eliminate $\dot\psi,\ \dot X^I,\ F_{0r}$ and
$\partial_\tau X^-$ in favor of the canonical variables and the
conjugate momenta. Let us start with $\partial_{\tau}X^-$ and recall
the definition of $\det N$ which is%
\be
 \det N=\det(N_{rs})(N_{00}-N_{0r}N^{rs}N_{s0}),
\ee%
where
\be\label{N00}
 N_{00}=-2\partial_{\tau}X^--\mu^2(X^I)^2+(\dot{X^I})^2+2i(\bar{\psi}\bar{\gamma}^-\partial_{\tau}\psi+\psi\bar{\gamma}^-\partial_{\tau}\bar{\psi})
 -4\mu\bar{\psi}\bar{\gamma}^-\Pi\psi
\ee%
and $N^{rs}$ is the inverse of $N_{rs}$, that is
$N^{rs}N_{sp}=\delta^r_p$. It is important to note that
$N^{00}\neq\frac{1}{N_{00}}$ because of
the off-diagonal electric-magnetic fields. By definition we have%
\be
 N^{00}=\frac{det({N_{rs})}}{detN}=\frac{det(g_{rs}+F_{rs})}{detN}
\ee%
The above two equations together with \eqref{p+Hlc} lead to %
\be\label{N001}%
 N_{00}=-(\frac{1}{p^+g_s})^2det(N_{rs})+N_{0r}N^{rs}N_{s0}
\ee%
Now we can eliminate $\partial_\tau X^-$  in \eqref{N00} using
\eqref{N001} and $\dot X^I$ using  \eqref{PI-def}. To eliminate
$F_{0r}$ for $P^r_E$ we use  \eqref{PrE-def} and recall that
\[N^{0r}=-\frac{\det(N_{rs})}{\det N}N^{sr}N_{0s},\]
yielding %
\be\label{c1}
 P^r_E=\frac{1}{g_s}\frac{\det (N_{rs})}{\sqrt{-\det
 N}}N^{sr}N_{0s}=p^+ N^{rs}N_{s0}.
\ee
In the above we have also used the following identities
\be\label{c2}\begin{split}%
 N^{0r}=&-N^{r0}\cr
N^{sr}N_{0s}=&-N^{rs}N_{s0},%
\end{split}\ee%
With the above the light-cone Hamiltonian \eqref{p+Hlc} is obtained to be%
\be\begin{split}
 P^-=&\frac{(P^I)^2}{2p^+}+\frac{(P^I_E)^2}{2p^+}+\frac{1}{2}\mu^2p^+(X^I)^2+\frac{1}{2p^+g_s^2}det(N_{rs})
 +2p^+\mu\bar{\psi}\bar{\gamma}^-\Pi\psi\cr
 +&\frac{\mu}{6 g_s}\epsilon^{rps}\bigl[\epsilon^{ijkl}
 X^i\partial_{r}X^j\partial_{p}X^k\partial_{s}X^l+\epsilon^{abcd}
 X^a\partial_{r}X^b\partial_{p}X^c\partial_{s}X^d\cr
 -&\partial_{r}X^I\partial_{p}X^J\bar{\psi}\gamma^{-IJ}\partial_{s}\psi\bigr]
\end{split}\ee%
where%
\be\label{PIE}%
 P^I_E=P^r_E\partial_r X^I.
\ee%
For the special case of D3-brane $det(F_{rs})=0$ and therefore%
\be
 \det(N_{rs})=det(g_{rs})+(B^I)^2,
\ee%
where%
\be
 B^I=B^r\partial_rX^I=-\frac{1}{\sqrt{2}}\epsilon^{rsp}\partial_rX^IF_{sp}.
\ee%
Putting all these together we find the light-cone Hamiltonian density%
\be\begin{split}
 H_{lc}=&\frac{(P^I)^2}{2p^+}+\frac{(P^I_E)^2}{2p^+}+\frac{1}{2}\mu^2p^+(X^I)^2+\frac{1}{2p^+g_s^2}det(g_{rs})
 +\frac{1}{2p^+g_s^2}(B^I)^2+2p^+\mu\bar{\psi}\bar{\gamma}^-\Pi\psi\cr
 +&\frac{\mu}{6 g_s}\big(\epsilon^{ijkl}
 X^i\{X^j, X^k, X^l\}+\epsilon^{abcd}
 X^a\{X^b, X^c, X^d\}\big)
 -\psi \gamma^{-IJ}\{X^I, X^J, \psi\},
\end{split}\ee%
which should be supplemented by \eqref{prE}.

  To obtain $H_{lc}$ given in
\eqref{Hlc} we need to move to the $SO(4)\times SO(4)$
representation for fermions. The details of which may be found in
Appendix C. Recall also that
\[\begin{split}
\{F, G, K\}&=\epsilon^{rps}\partial_r F\ \partial_p G\ \partial_s
K \cr \det(g_{rs})&= \det(\partial_r X^I\partial_s
X^I)=\frac{1}{3!}\{ X^I, X^J, X^K\}\{X^I, X^J,X^K\}.
\end{split}\]
%
%

\section{Fermionic Notations}
\subsection{Metsaev's fermionic notation}\label{Metsaev-fermion}
For completeness we summarize the fermionic notation which is used
in section 2.1 \cite{0211178}. Chiral representation are used for
$32\times32$ matrices $\Gamma$ in terms of $16\times16$ matrices
$\gamma$
\begin{displaymath}%
 \Gamma^\mu=\left(\begin{array}{cc}
 0 &\gamma^\mu \\
 \bar{\gamma}^\mu & 0 \\
\end{array}\right),\ \ \ \mu=0,1,...,9
\end{displaymath}%
\be%
 \gamma^\mu\bar{\gamma^\nu}+\gamma^\nu\bar{\gamma}^\mu=2\eta^{\mu\nu}\ \ ,\ \
 \gamma^\mu=(\gamma^\mu)^{\alpha\beta}\ \ ,\ \
 \bar{\gamma}^\mu=(\bar{\gamma}^\mu)_{\alpha\beta}%
\ee%
\be%
 \gamma^\mu=(\textbf{1},\gamma^I,\gamma^9)\ \ ,\ \
 \bar{\gamma}^\mu=(-\textbf{1},\gamma^I,\gamma^9)\ \ ,\ \
 \alpha,\beta=1,...,16%
\ee%
note that all $\gamma^\mu$ matrices are real and symmetric
and %
\be\begin{split}%
 (\gamma^{\mu\nu})^{\ \alpha}_\beta&\equiv\frac{1}{2}(\gamma^\mu\bar{\gamma}^\nu)^{\ \alpha}_\beta-(\mu\leftrightarrow\nu)\cr
 (\bar{\gamma}^{\mu\nu})_\alpha^{\ \beta}&\equiv\frac{1}{2}(\bar{\gamma}^\mu\gamma^\nu)_\alpha^{\ \beta}-(\mu\leftrightarrow\nu)\cr
 \gamma^{+-}&=\gamma^0\gamma^9%
\end{split}\ee%
The 32-component positive and negative chirality spinor are
decomposed in term of 16-component spinors as%
\begin{displaymath}%
 \psi=\left(\begin{array}{c}
 \psi^\alpha\\
 0\\
\end{array}\right)\ \ ,\ \ %
 \theta=\left(\begin{array}{c}
 0 \\
 \theta_\alpha \\
\end{array}\right)%
\end{displaymath}%
The complex Weyl spinor $\psi$ is related to two real Majorana-Weyl
spinors $\psi^1$ and $\psi^2$ by%
\be %
 \psi=\frac{1}{\sqrt{2}}(\psi^1+i\psi^2)\ \ \ ,\ \ \
 \bar{\psi}=\frac{1}{\sqrt{2}}(\psi^1-i\psi^2)
\ee%
The short-hand notation  $\bar{\psi}\gamma^\mu\psi$ stands for
$\bar{\psi}^\alpha\gamma^\mu_{\alpha\beta}\psi^\beta$ and alike for
similar bi-fermions.
%
%
\subsection{The $SO(4)\times SO(4)$ fermionic
notation}\label{fermion-so4so4}
The Dirac matrices in ten dimensions obey%
\be
 \{\Gamma^\mu,\Gamma^\nu\}=2g^{\mu\nu}
\ee%
A convenient choice of basis for $32\times32$ matrices, which is
denoted by $\Gamma$, is in term of $16\times16$ matrices
$\gamma$%
\begin{displaymath}%
 \Gamma^+=i\left(\begin{array}{cc}
 0 &\sqrt{2} \\
 0 & 0 \\
\end{array}\right),\ %
\Gamma^-=i\left(\begin{array}{cc}
 0& 0 \\
 \sqrt{2}& 0 \\
\end{array}\right),\ %
 \Gamma^I=\left(\begin{array}{cc}
 \gamma^I& 0 \\
 0& -\gamma^I \\
\end{array}\right),\ %
 \Gamma^{11}=\left(\begin{array}{cc}
 \gamma^{(8)}& 0 \\
 0& -\gamma^{(8)}\\
\end{array}\right)
\end{displaymath}%
and the $\gamma$ satisfy $\{\gamma^I,\gamma^J\}=2\delta^{IJ}$ with
$\delta^{IJ}$ the metric on the transverse space and
$\Gamma^\pm=\frac{1}{\sqrt{2}}(\Gamma^0\pm\Gamma^9)$. We may choose
our ten dimensional, 32 components Majorana fermions $\psi$ to
satisfy%
\be%
 \Gamma^+\psi^+=0\ \ ,\ \ \Gamma^-\psi^-=0
\ee%
and it is easily seen that %
\begin{displaymath}%
 \psi^+=\left(\begin{array}{c}
 \psi^+_\alpha \\
 0\\
\end{array}\right),\ %
 \psi^-=\left(\begin{array}{c}
 0 \\
 \psi^-_\alpha \\
\end{array}\right),\ \alpha=1,...,16.%
\end{displaymath}%
where $\psi^\pm_\alpha$ can be thought as $SO(8)$ Majorana fermions
and the $\gamma^I$ matrices as $16\times16$ $SO(8)$ Majorana gamma
matrices. Moreover, we have%
\begin{displaymath}%
 \Gamma^{11}\psi^+=\left(\begin{array}{c}
 \gamma^{(8)}\psi^+_\alpha \\
 0\\
\end{array}\right),\ %
 \Gamma^{11}\psi^-=\left(\begin{array}{c}
 0 \\
 -\gamma^{(8)}\psi^-_\alpha \\
\end{array}\right)%
\end{displaymath}%
i.e. the ten dimensional chirality is related to eight dimensional
$SO(8)$ chirality.

 Let us now consider type IIB theory on the plane-wave
background. In this case we start with ten dimensional fermions of
the same chirality. As stated in the above equations,
$\psi^\pm_\alpha$ should have $\pm$
$SO(8)$ chirality i.e. %
\be%
 (\gamma^{(8)}\psi^\pm)_\alpha=\pm\psi^\pm_\alpha
\ee%
By selecting $\gamma^{(8)}=diag(\textbf{1}_{8},-\textbf{1}_{8})$ the
last equation is easily solved and hence%
\begin{displaymath}%
 \psi^+_\alpha=\left(\begin{array}{c}
 \psi^+_a \\
 0\\
\end{array}\right),\ %
 \psi^-=\left(\begin{array}{c}
 0 \\
 \psi^-_{\dot{a}} \\
\end{array}\right),\ a,\dot{a}=1,...,8.%
\end{displaymath}%
where $\psi^+_a$,$\psi^-_{\dot{a}}$ are Weyl-Majorana fermions. The
gamma matrices can also be reduced to $8\times8$ representation,
$\gamma^I_{a\dot{a}}$ and $\gamma^I_{\dot{a}a}$,
\begin{displaymath}%
 \gamma^I=\left(\begin{array}{cc}
 0 & \gamma^I_{a\dot{a}}\\
 \gamma^I_{\dot{a}a} & 0\\
\end{array}\right),\ \ a,\dot{a},I=1,...,8.%
\end{displaymath}\\%
In the plane-wave background, due to the presence of RR form flux,
the $SO(8)$ is broken  to $SO(4) \times SO(4)$. It is therefore
better to adopt $SO(4) \times SO(4)$ representation for fermions.
$SO(4)$ Dirac fermion can be decomposed into two Weyl fermions
$\psi_{\alpha}$ and $\psi_{\dot{\alpha}}, \ \alpha,
\dot{\alpha}=1,2$. As usual for $SU(2)$ fermion Weyl indices are
lowered and raised by $\epsilon$ tensor%
\be%
 \psi_{\alpha}=\epsilon_{\alpha\beta}\psi^\beta%
\ee%
Hence our fermions have two indices which are labeled to two
different $SO(4)$ Weyl indices i.e.%
\be\begin{split}%
 \psi_a\rightarrow\psi_{\alpha\beta}\ , \psi_{\dot{\alpha}\dot{\beta}}\cr%
 \psi_{\dot{a}}\rightarrow\psi_{\dot{\alpha}\beta}\ ,  \psi_{\alpha\dot{\beta}}%
\end{split}\ee%
where the first (second) indices are related to first (second)
$SO(4)$. The Weyl indices are lowered and raised by two $\epsilon$ tensor%
\be
 \psi_{\alpha\beta}^{\dagger}\equiv\epsilon_{\alpha\rho}\epsilon_{\beta\lambda}\psi^{\dagger\rho\lambda}
 \ \ \ ,\ \ \
 \psi_{\dot{\alpha}\dot{\beta}}^{\dagger}\equiv\epsilon_{\dot{\alpha}\dot{\rho}}\epsilon_{\dot{\beta}\dot{\lambda}}\psi^{\dagger\dot{\rho}\dot{\lambda}}
\ee%
where%
\be
 (\psi_{\alpha\beta})^\dagger=\psi^{\dagger\alpha\beta}\ \ \ ,\ \
 \
 (\psi_{\dot{\alpha}\dot{\beta}})^\dagger=\psi^{\dagger\dot{\alpha}\dot{\beta}}
\ee%
\be
 (\psi^{\dagger}_{\alpha\beta})^\dagger=\psi^{\alpha\beta}\ \ \
 ,\ \ \
 (\psi^{\dagger}_{\dot{\alpha}\dot{\beta}})^\dagger=\psi^{\dot{\alpha}\dot{\beta}}
\ee%
We also choose a proper basis for $\gamma^I_{a\dot{a}}$
which is %
\be%
 \gamma^I_{a\dot{a}}=(\gamma^i_{a\dot{a}},\gamma^a_{a\dot{a}})%
\ee%
where%
\begin{displaymath}%
 \gamma^i_{a\dot{a}}= \left(\begin{array}{cc}
 0 & (\sigma^i)_{\alpha\dot{\beta}}\delta_\rho^\theta\\
 (\sigma^i)^{\dot{\alpha}\beta}\delta_{\dot{\rho}}^{\dot{\theta}} & 0 \\
\end{array}\right)\ \ ,\ \ %
 \gamma^i_{\dot{a}a}= \left(\begin{array}{cc}
 0 & (\sigma^i)_{\alpha\dot{\beta}}\delta_{\dot{\rho}}^{\dot{\theta}}\\
 (\sigma^i)^{\dot{\alpha}\beta}\delta_\rho^\theta & 0 \\
\end{array}\right)%
\end{displaymath}%
and%
\begin{displaymath}%
 \gamma^a_{a\dot{a}}= \left(\begin{array}{cc}
  -\delta_\alpha^\beta(\sigma^a)_{\rho\dot{\theta}} & 0\\
 0 & \delta_{\dot{\alpha}}^{\dot{\beta}}(\sigma^a)_{\dot{\rho}\theta} \\
\end{array}\right)\ \ ,\ \ %
 \gamma^a_{\dot{a}a}= \left(\begin{array}{cc}
  -\delta_\alpha^\beta(\sigma^a)^{\dot{\rho}\theta} & 0\\
 0 & \delta_{\dot{\alpha}}^{\dot{\beta}}(\sigma^a)^{\rho\dot{\theta}} \\
\end{array}\right)%
\end{displaymath}%
with%
\be%
(\sigma^i)_{\alpha\dot{\alpha}}=(\sigma^1,\sigma^2,\sigma^3,\textbf{1})_{\alpha\dot{\alpha}}%
\ee

In performing the superalgebra analysis we have used the following
$\sigma$ matrix identities
\be%
(\sigma^a)_{\alpha\dot{\alpha}}=(\sigma^1,\sigma^2,\sigma^3,\textbf{1})_{\alpha\dot{\alpha}}%
\ee%
\bse\begin{align} %
 \big((\sigma^{i})_\alpha^{\ \ \dot{\beta}}\big)^\dagger=(\sigma^{i})_{\dot{\beta}}^{\ \ \alpha}\ \ &,\ \ %
 \big((\sigma^{a})_\alpha^{\ \ \dot{\beta}}\big)^\dagger=(\sigma^{a})_{\dot{\beta}}^{\ \ \alpha}\\ %
 \big((\sigma^{i})_{\dot{\alpha}}^{\ \ \beta}\big)^\dagger=(\sigma^{i})_{\beta}^{\ \ \dot{\alpha}}\ \ &,\ \ %
 \big((\sigma^{a})_{\dot{\alpha}}^{\ \ \beta}\big)^\dagger=(\sigma^{a})_{\beta}^{\ \ \dot{\alpha}}\\ %
 \big((\sigma^{ij})_{\alpha}^{\ \ \beta}\big)^\dagger=-(\sigma^{ij})_{\beta}^{\ \ \alpha}\ \ &,\ \ %
 \big((\sigma^{ab})_{\alpha}^{\ \ \beta}\big)^\dagger=-(\sigma^{ab})_{\beta}^{\ \ \alpha}\\ %
 \big((\sigma^{ij})_{\dot{\alpha}}^{\ \ \dot{\beta}}\big)^\dagger=-(\sigma^{ij})_{\dot{\beta}}^{\ \ \dot{\alpha}}\ \ &,\ \ %
 \big((\sigma^{ab})_{\dot{\alpha}}^{\ \ \dot{\beta}}\big)^\dagger=-(\sigma^{ab})_{\dot{\beta}}^{\ \ \dot{\alpha}} \\%
 (\sigma^{i})_\alpha^{\ \ \dot{\beta}}=\epsilon_{\alpha\rho}(\sigma^{i})^{\rho\dot{\beta}}=(\sigma^{i})_{\alpha\dot{\rho}}\epsilon^{\dot{\rho}\dot{\beta}}\ \ &,\ \ %
 (\sigma^{a})_\alpha^{\ \ \dot{\beta}}=\epsilon_{\alpha\rho}(\sigma^{a})^{\rho\dot{\beta}}=(\sigma^{a})_{\alpha\dot{\rho}}\epsilon^{\dot{\rho}\dot{\beta}}\\ %
 (\sigma^{ij})_\alpha^{\ \ \beta}=\epsilon_{\alpha\rho}(\sigma^{ij})^{\rho\beta}=(\sigma^{ij})_{\alpha\rho}\epsilon^{\rho\beta}\ \ &,\ \ %
 (\sigma^{ab})_\alpha^{\ \ \beta}=\epsilon_{\alpha\rho}(\sigma^{ab})^{\rho\beta}=(\sigma^{ab})_{\alpha\rho}\epsilon^{\rho\beta} %
\end{align}\ese %
Additional discussion about notation can be found
in \cite{0310119}.%
%

%
%

\section{BPS equation}\label{BPS-Appendix}%

The BPS equations \eqref{BPS-static} are equations relating
$\epsilon$ and $\epsilon^\dagger$. Let us first focus on
(\ref{BPS-static}a), which can be written as%
\be\label{BPS-eq}%
(\tilde X^i \sigma^i)^{\dot\alpha}_{\
\rho}\epsilon_{\dot\alpha\lambda}=(\tilde \Pi^i
\sigma^i)^{\dot\alpha}_{\ \rho}\epsilon^\dagger_{\dot\alpha\lambda},
\ee%
where
\[
\Pi^i=\frac{B^i}{g_s}+i P^i_E\ .
\]
Next we note that%
\be\label{X2}%
(\tilde X^i \sigma^i)^{\dot\alpha}_{\ \rho} (\tilde X^j \sigma^j)^{
\rho}_{\ \dot\beta}=|\tilde X|^2 \delta^{\dot\alpha}_{\ \dot\beta}
\ee%
and
\be\label{Pi2}%
( \Pi^i \sigma^i)^{\dot\alpha}_{\ \rho} ( \bar \Pi^j \sigma^j)^{
\rho}_{\ \dot\beta}=\Pi^i\bar\Pi^i \delta_{\dot\alpha}^{\
\dot\beta}+\Pi^i\bar\Pi^j(\sigma^{ij})^{\dot\alpha}_{\ \dot\beta}.
\ee%
Using the above two identities \eqref{BPS-eq} leads to
\be\label{epsReps} %
 \epsilon_{\dot{\beta}\lambda}={\cal{R}}_{\ \dot{\beta}}^{\dot{\alpha}}\epsilon^\dagger_{\dot{\alpha}\lambda}=
 \frac{\tilde{X}^j\Pi^i}{|\tilde{X}|^2}\big((\sigma^{ij})^{\dot{\alpha}}_{\ \dot{\beta}}
 +\delta^{ij}\delta^{\dot{\alpha}}_{\
 \dot{\beta}}\big)\epsilon^\dagger_{\dot{\alpha}\lambda}\ ,
\ee %
and
\be\label{epsilon1} %
 \tilde{X^i}\bp^j\big((\sigma^{ij})^{\dot{\alpha}}_{\ \dot{\beta}}+\delta^{ij}\delta^{\dot{\alpha}}_{\ \dot{\beta}}\big)\epsilon_{\dot{\alpha}\lambda}=
 \Pi^i\bp^j\big((\sigma^{ij})^{\dot{\alpha}}_{\ \dot{\beta}}+\delta^{ij}\delta^{\dot{\alpha}}_{\
 \dot{\beta}}\big)\epsilon^\dagger_{\dot{\alpha}\lambda}\ .
\ee %
Taking the complex conjugate of \eqref{epsReps},%
\be\label{epsilon2} %
 \epsilon^\dagger_{\dot{\beta}\lambda}=
 \frac{\tilde{X}^i\bp^j}{|\tilde{X}|^2}\big((\sigma^{ij})^{\dot{\alpha}}_{\ \dot{\beta}}
 +\delta^{ij}\delta^{\dot{\alpha}}_{\ \dot{\beta}}\big)\epsilon_{\dot{\alpha}\lambda}
\ee %
and substituting the R.H.S. of \eqref{epsilon2} into the L.H.S of \eqref{epsilon1} we obtain%
\be\begin{split} %
 \bigg((|\tilde{X}|^2-|\Pi|^2)\delta^{\dot{\alpha}}_{\
 \dot{\beta}}-\Pi^i\bp^j(\sigma^{ij})^{\dot{\alpha}}_{\
 \dot{\beta}}\bigg)\epsilon^\dagger_{\dot{\alpha}\lambda}={\cal M}_{\
 \dot{\beta}}^{\dot{\alpha}}\epsilon^\dagger_{\dot{\alpha}\lambda}=0
\end{split}\ee %
where $|\Pi|^2=\Pi^i\bp^i$. One can exactly follow the same computations for (\ref{BPS-static}b) to obtain%
\[ %
 \epsilon_{\beta\dot{\lambda}}={\cal{R}}_{\ \beta}^{\alpha}\epsilon^\dagger_{\alpha\dot{\lambda}}=
 \frac{\tilde{X}^j\Pi^i}{|\tilde{X}|^2}\big((\sigma^{ij})^{\alpha}_{\ \ \beta}
 +\delta^{ij}\delta^{\alpha}_{\ \beta}\big)\epsilon^\dagger_{\alpha\dot{\lambda}}
\] %
and then%
\be\begin{split} %
 \bigg((|\tilde{X}|^2-|\Pi|^2)\delta^{\alpha}_{\
 \beta}-\Pi^i\bp^j(\sigma^{ij})^{\alpha}_{\
 \beta}\bigg)\epsilon^\dagger_{\alpha\dot{\lambda}}={\cal M}_{\
 \beta}^{\alpha}\epsilon^\dagger_{\alpha\dot{\lambda}}=0
\end{split}\ee %
${\cal M}$ as a $2\times 2$ matrix can have one or two zero
eigenvalue which support $1/16$ or $1/8$ BPS configurations for
each of the two equations. Therefore, we can have
 $1/4$ or $1/8$ BPS configurations.\\ %
$\bullet\ \ 1/4\ BPS\  configurations$%

The first one takes place when ${\cal M}_{\
 \dot{\beta}}^{\dot{\alpha}}\equiv0$ and then%
\bse\begin{align} %
 |\tilde{X}|^2-|\Pi|^2&=0\\
 \Pi^{[i}\bp^{j]}&=0 %
\end{align}\ese %
Thus BPS equations have three different options:%

 {\it i)}$\ \ \ P_E\neq0,B=0\rightarrow|\tilde{X}|^2=|P_E|^2$%

 {\it ii)}$\ \ P_E=0,B\neq0\rightarrow|\tilde{X}|^2=\frac{1}{g^2_s}|B|^2$%

{\it iii)}$\  P_E\neq0,B\neq0\rightarrow|\tilde{X}|^2=|\bp|^2\ \ provided\ that\ \ P^i_EB^j=P^j_EB^i$\\%
The third case corresponds to parallel electric and magnetic fields because %
\be %
B^j=\frac{B\cdot P_E}{P_E^2}P_E^j\ . %
\ee%
\\
$\bullet\ \ 1/8\ BPS\  configurations$ %

If ${\cal M}$ has only one zero eigenvalue we obtain $1/8$ BPS
configuration. In order ${\cal M}$ to have one zero eigenvalue
\be %
 (|\tilde{X}|^2-|\Pi|^2)^2+\Pi^i\bp^j\Pi^{[i}\bp^{j]}=0.%
\ee %
It is obvious that when $\Pi^i\|\bp^i$ the second term vanishes and
then this equation recover $1/4$ BPS equation.\\
%
%

%

\end{document}